\documentstyle[epsfig, 12pt]{article}

\def\dalemb#1#2{{\vbox{\hrule height .#2pt
        \hbox{\vrule width.#2pt height#1pt \kern#1pt
                \vrule width.#2pt}
        \hrule height.#2pt}}}

\let\a=\alpha  \let\g=\gamma \let\d=\delta \let\e=\epsilon
\let\z=\zeta  \let\th=\theta  \let\k=\kappa
\let\l=\lambda  \let\n=\nu \let\x=\xi  
\let\s=\sigma     
 
        \let\Th=\Theta 
\let\X=\Xi  \let\S=\Sigma  \let\Y=\Psi
 
\let\la=\label  
  
\def\nn{\nonumber} \def\bd{\begin{document}} \def\ed{\end{document}}
\def\ds{\documentstyle} \let\fr=\frac \let\bl=\bigl \let\br=\bigr
\let\Br=\Bigr \let\Bl=\Bigl
\let\bm=\bibitem
\let\na=\nabla
\def\tU{{\widetilde U}}
\let\pa=\partial \let\ov=\overline
\def\ie{{\it i.e.\ }}
\newcommand{\be}{\begin{equation}}
\newcommand{\ee}{\end{equation}}
\def\ba{\begin{array}}
\def\ea{\end{array}}
\def\ft#1#2{{\textstyle{{\scriptstyle #1}\over {\scriptstyle #2}}}}
\def\fft#1#2{{#1 \over #2}}
\def\F#1#2{{ F_{#1}^{(#2)} }}
\def\cF#1#2{{ {\cal F}_{#1}^{(#2)} }}

\def\R{{\bf R}}
\def\sst#1{{\scriptscriptstyle #1}}
\def\oneone{\rlap 1\mkern4mu{\rm l}}
\def\e7{E_{7(+7)}}
\def\td{\tilde}
\def\wtd{\widetilde}
\def\im{{\rm i}}
\def\bog{Bogomol'nyi\ }
\newcommand{\ho}[1]{$\, ^{#1}$}
\newcommand{\hoch}[1]{$\, ^{#1}$}
\newcommand{\bea}{\begin{eqnarray}}
\newcommand{\eea}{\end{eqnarray}}
\newcommand{\ra}{\rightarrow}
\newcommand{\lra}{\longrightarrow}
\newcommand{\Lra}{\Leftrightarrow}
\newcommand{\ap}{\alpha^\prime}
\newcommand{\bp}{\tilde \beta^\prime}
\newcommand{\cB}{{\cal B}}
\newcommand{\cO}{{\cal O}}
\newcommand{\vecx}{\vec{x}}
\newcommand{\vecy}{\vec{y}}
\newcommand{\vecp}{\vec{p}}
\newcommand{\vecq}{\vec{q}}
\newcommand{\tr}{{\rm tr} }
\newcommand{\Tr}{{\rm Tr} }
\newcommand{\NP}{Nucl. Phys. }

\newcommand{\cL}{{\cal L}}
\newcommand{\cA}{{\cal A}}
\newcommand{\cD}{{\cal D}}

\def\sst#1{{\scriptscriptstyle #1}}
\def\0{{\sst{(0)}}}
\def\1{{\sst{(1)}}}
\def\2{{\sst{(2)}}}
\def\3{{\sst{(3)}}}
\def\4{{\sst{(4)}}}
\def\5{{\sst{(5)}}}
\def\6{{\sst{(6)}}}
\def\7{{\sst{(7)}}}
\def\8{{\sst{(8)}}}
\def\ve{\varepsilon}
\def\vf{\varphi}
\def\F{\Phi}
\def\wg{\wedge}

\newcommand{\tamphys}{\it 
}

\newcommand{\auth}{AUTHORS}

\def\thb{\bar{\theta}}
\def\Thb{\bar{\Theta}}
\def\barp{\bar{p}}
\def\barq{\bar{q}}
\def\barc{\bar{c}}
\def\bard{\bar{d}}
\def\e{\epsilon}

\def \bi{\bibitem}
\def \la {\label}

\def \l {\lambda}
\def\foot{\footnote}
\def \tl  {{\tilde \l}}
\def \sql {{\sqrt \l}}
\def \adss {$AdS_5 \times S^5$\ }
\newcommand{\rf}[1]{(\ref{#1})}
\def \ov {\over}

\def\th{\theta}
\def\Th{\Theta}
\def\vth{\vartheta}
\def\btheta{{\bar\theta}}
\def\ttheta{{{\tilde\theta}}}
\def\bttheta{{{\bar\ttheta}}}
\def\vth{\vartheta}

\def\ra{\rightarrow}
\def\N{{\cal N}}
\def\F{{\cal F}}
\def\uM{\underline{M}}
\def\uN{\underline{N}}
\def\uP{\underline{P}}
\def\cc{\circ}
\def\eqv{\equiv}

\def\ni{\noindent}

\def\Ep{E^{{}^{(+)}}}
\def\Em{E^{{}^{(-)}}}

\def\Mp{M^{{}^{(+)}}}
\def\Mm{M^{{}^{(-)}}}

\def \ha{{1\ov 2}}

\def\r{\rho}

\def\Y{{\rm Y}}
\def\X{{\rm X}}
\def\tY{\tilde{\rm Y}}
\def\tX{\tilde{\rm X}}
\def\dY{\dot{\rm Y}}
\def\dX{\dot{\rm X}}

\def \J {\mathcal{J}}
\def \del {\partial}

\def\dF{\dot{F}}
\def\dG{\dot{G}}
\def\df{\dot{f}}
\def \E {{\cal E}}
\def \S {{\cal S}}
\def \J {{\cal J}}

\def\ms{\mathcal{S}}
\def\mj{\mathcal{J}}
\def\soj{\fr{\ms}{\mj}}
\def \R {{\bf R}}
\def \om {\omega}
\def \bE {\bar E}
\def \x {{\cal X}}

\def \bi{\bibitem}
\def \la {\label}

\def \l {\lambda}
\def\foot{\footnote}
\def \tl  {{\tilde \l}}
\def \sql {{\sqrt \l}}
\def \adss {$AdS_5 \times S^5$\ }
\def \ov {\over}

\def \varpi {{\rm w}}

\def\thb{\bar{\theta}}
\def\Thb{\bar{\Theta}}
\def\zb{\bar{z}}
\def\psib{\bar{\psi}}
\def\barp{\bar{p}}
\def\barq{\bar{q}}
\def\barc{\bar{c}}
\def\bard{\bar{d}}
\def\e{\epsilon}
\def\wb{\bar{w}}
\def\lb{\bar{\l}}
\def\Jb{\bar{J}}
\def\Nb{\bar{N}}

\def\At{\tilde{A}}
\def\Bt{\tilde{B}}
\def\Ct{\tilde{C}}
\def\Dt{\tilde{D}}
\def\Et{\tilde{E}}
\def\Ft{\tilde{F}}
\def\Gt{\tilde{G}}
\def\Mt{\tilde{M}}
\def\at{\tilde{a}}
\def\bt{\tilde{b}}
\def\ct{\tilde{c}}
\def\dt{\tilde{d}}
\def\et{\tilde{e}}
\def\ft{\tilde{f}}
\def\gt{\tilde{g}}
\def\ola{\overleftarrow}
\def\ora{\overrightarrow}
\def\at{\tilde{\a}}

\def\ps{\rlap{\, /}\;\,p }
\def\ks{\rlap{\, /}\;\,k }

\def\gym{g_{YM}}

\def\adot{\dot{a}}
\def\bdot{\dot{b}}
\def\bpa{\bar{\pa}}

\newcommand{\PL}{{\em Phys.\ Lett.\ }}
\newcommand{\PR}{{\em Phys.\ Rev.\ }}
\newcommand{\PRP}{{\em Phys.\ Rep.\ }}
\newcommand{\CMP}{{\em Comm.\ Math.\ Phys.\ }}
\newcommand{\MPL}{{\em Mod.\ Phys.\ Lett.\ }}
\newcommand{\PRL}{{\em Phys.\ Rev.\ Lett.\ }}
\newcommand{\IJMP}{{\em Int.\ J.\ Mod.\ Phys.\ }}

\begin{document}
\overfullrule=0pt
\parskip=2pt
\parindent=12pt
\headheight=0in \headsep=0in \topmargin=0in \oddsidemargin=0in

\vspace{ -3cm} \thispagestyle{empty}

\begin{center}

{\Large\bf  Geometric counter-vertex for open string scattering on
D-branes
  }

 \vspace{.5cm} { I.Y. Park  }\\
 \vskip 0.2cm

{\it Kavli Institute for Theoretical Physics,\\
 Santa Barbara, California, USA \\
} \vspace{0.3cm}

{\it Department of Chemistry and Physics, University of Arkansas at Pine Bluff\\
Pine Bluff, AR 71601, USA} \\

 \vspace{0.3cm}
 and\\
 \vspace{0.3cm}

{\it Division of Natural and Physical Sciences, Philander Smith College\\
Little Rock, AR 72202, USA \\
inyongpark05@gmail.com}

\end{center}

 \vspace{0.1cm}

 \begin{abstract}
\ni In arXiv:0801.0218 [hep-th] it was conjectured that quantum
effects of open strings moving on D-branes generate the D-brane
geometry through a counter-vertex operator. The conjecture has been
checked at one-loop in arXiv:0806.3330 [hep-th]. Here we discuss the
two-loop extension.

\end{abstract}
\newpage

\setcounter{equation}{0} \setcounter{footnote}{0}
\setcounter{section}{0}


\section{Introduction}
Before
 the birth of the D-brane physics \cite{Polchinski:1995mt,pol},
 an open string was a rather subsidiary
 object with a closed string at the center of attention, both in
 theory and phenomenology. The advent of the D-brane physics has brought
  a shift in importance toward
 the open string: the open string has become to play much more important
 roles in string theory.
 What makes it tempting to consider a further shift is
 the potential new physics that is associated with the end points of an
 open string which might have not been fully appreciated. On a corner
 of the moduli space the end points of an open string may
 stick together, thereby converting it into a closed
 string \cite{Nielsen:1973qs,Gibbons:2000hf,Sen:2000kd,Park:2001bm}.
 This seems to suggest unification of an open string and a closed string
  at the level of degrees of freedom. It may be
 technically challenging but should be a worthwhile endeavor to try to
 realize a closed string as some sort of composite or bound state of an open string
  in the frame-work of purely open string degrees of freedom.

 An additional motivation for attempting a further shift comes from
 the recent works \cite{Park:2007mc,Park:2008sg,Park:2008fp} where
it was conjectured that quantum effects of open strings moving on
D-branes generate the D-brane geometry through a counter-vertex
operator.  The divergence structures
 were analyzed for the scattering amplitudes of massless open strings.
 (Previous works on scattering that involve a D-brane
 can be found e.g., in \cite{Hashimoto:1996bf}. Also the works of
 \cite{Garousi:2007fk,Pesando:1999hm} are more closely related to \cite{Park:2007mc}.\footnote{We thank
 Ehsan Hatefi for bringing the papers in \cite{Garousi:2007fk} to our attention.}) For the analysis, first
 the massless vertex operators were constructed for
 the external open strings on the D-branes. One thing note-worthy is that the
 closure under super symmetry transformation requires
 the momenta of the scattering states to be solely along the brane
 directions. (It should be possible to understand this on more physical grounds.)
 Because of this constraint, one expects that
 the loop effects and the analysis thereof will go differently from those of
 a D9-brane since parts of the momentum degrees of freedom are lost.
 Indeed, that is what we have observed.
 It was anticipated that it should be possible to remove the
 divergences by a composite operator.  The composite operator is, in essence,
 the non-linear sigma model action for the corresponding D-brane
 geometry \cite{Cvetic:1999zs}\cite{Sahakian:2004gy}\cite{Mizoguchi:2002qy}.
 \footnote{In a broader sense, a
 connection between the loop effects and geometry goes back to the
 Fischler-Susskind mechanism \cite{Fischler:1986ci}. Also the result
 of \cite{Gonzalez-Rey:1998uh} can be interpreted in the context of the
 present paper.
  For the scattering of {\em massless} states the D-branes geometry
 is that of the extremal case. At least that is what the results so far
 indicate.
 } The anticipation was verified at one-loop in the
 subsequent work \cite{Park:2008fp}. In this work, we set out to establish
  the two-loop extension of the conjecture.

The task consists of two parts. First one must obtain the two-loop
divergence structures. With those available one can carry out a
check whether the counter vertex produces suitable amplitudes to
cancel the divergences. The two-loop amplitudes were fully
calculated only recently. It was first done in the RNS formulation
\cite{Zheng:2002ji}\cite{D'Hoker:2005jc,D'Hoker:2001nj} and later in
the pure spinor formulation \cite{Berkovits:2005df}. (For the
bosonic case see also \cite{DiVecchia:1996uq,DiVecchia:1988cy}.) The
connection with the geometry can, a priori, be such that the counter
vertex operator may depend on the scattering states and the number
of loops. However, the results so far indicate that things may work
much more remarkable way: the counter vertex operator given in
(\ref{startingaction5quoted'}) may be the {\em master} operator
 that accounts for any and all the scattering amplitudes, at least for the
 massless cases.

We would like to make a few remarks before getting into the detailed
analysis. Firstly, compared with the previous work, the new
ingredient for the two-loop analysis is an introduction of the
multiplicative renormalization of the external vertex operators.
 We also introduce the string
tension renormalization for future purposes though it is not
necessary for the amplitudes that we consider at the given order,
i.e., the two-loop four point vector and scalar amplitude. Once one
considers various other amplitudes, which is one of the near-future
tasks that we will work on, one may need such renormalization. The
second remark is that we minimize the use of T-duality but rather
rely on direct computations. That is because we believe that there
are subtle issues with T-duality. The issues may have something to
do with the fact that the open string states on the D-branes do not
have transverse components of their momenta. We postpone the
discussion on this until the conclusion. Lastly, when we consider an
open string on a Dp brane with $p<9$, it seems that one doesn't have
to consider the non-planar graphs. For a non-planar graph at least
one of the external state must momentarily leave the brane, which
would violate the boundary conditions. There will be
more on this in the conclusion.\\

The organization of the paper is as follows. In sec2, we explain the
strategy and summarize the results. After a brief review of the
one-loop divergence cancellation, we write down and examine the
two-loop four point open string amplitude. It can be obtained by
inspecting the amplitude for the closed string which has been
obtained in the literature in different formulations. It is noted
that the (most) divergent part of the amplitude comes with an
overall factor of $s+t+u$ where $s,t,u$ are the Mandelstam
variables, hence vanishes. Therefore it seems that the two-loop
counter vertex operator should not generate any divergence since
there is no divergence to cancel against in the first place. In
sec3, we confirm this by explicit computations up to the following
point. It turns out that some of the vertices in the counter vertex
operator do produce divergences. However, the forms of the
divergences are such that they have precisely the same forms as the
tree level ones. That makes it possible for them to be absorbed by
renormalization of the external vertex operators,\footnote{The
renormalization of the external vertex seems in the same spirit in
the work by S. Weinberg, \cite{Weinberg:1985je}.} as shown in sec2.
In sec3, we verify by explicit computations that the forms of the
divergences are the same as those of the tree level ones. In sec4,
we discuss various issues that include some three-loop perspectives
and future directions.

\section{Strategy and summary of results}

We start with a brief review of the one-loop case. As in the
previous works \cite{Park:2007mc,Park:2008sg,Park:2008fp}, we use
the light-cone formulation. More details of the conventions and some
simpler calculations can be found in the same references. As
well-known the formulation is not suited for two- or higher loop
computations. For that, one needs to use the RNS formulation
\cite{D'Hoker:2005jc} or the pure spinor formulation
\cite{Berkovits:2005df}. On the contrary, it is well-suited for our
purpose which is to compute the {\em tree level} correlators with
the counter vertex operator inserted. It is rather awkward that we
need to employ more than one formulation: one for computing the tree
diagrams with counter vertex inserted and another for computing the
loop diagrams. The benefit is that it is presumably the easiest in
light-cone formulation to write down the form of the counter vertex
operator. It is also straightforward to compute various tree level
correlators with counter vertices inserted.

In the one-loop case one can use the light-cone formulation, both
for computing the one-loop diagram, $<VVVV>_{\mbox{1 loop}}$, and
the tree diagram with the counter vertex operator, $V_G$, inserted,
$<VVVV\;V_G>_{\mbox{tree}}$. The subscript $G$ indicates its
connection to geometry. The geometry vertex operator
\cite{Park:2008fp} before any expansion\footnote{As commented in
some of the previous works, e.g.,\cite{Park:2008fp}, a correlator
with the series inserted is likely to terminate after some orders in
an expansion that we call a large-$r_0$ expansion
\cite{Park:2008sg,Park:2008fp}. For the scalar scattering, it does
terminate due to dimensional regularization. For the vector
scattering there is a general tendency that a correlator with many
fields inserted vanishes due to one or more reasons presented right
above sec3.1.} is given by
 \bea
 V_G
 =&&\int  \;(-)\fr{1}2\sqrt{h}\;h^{ij}
 \left[\pa_i X^u \pa_j X^v \eta_{uv} (H^{-1/2}-1)+
 \pa_i X^m \pa_j X^n \eta_{mn} (H^{1/2}-1)\right]
  \nn\\
 && +\fr1{2p^+} \left[\fr{}{}\right.
   -2i(\sqrt{h}\;h^{ij}-\ve^{ij})\pa_i X^+ H^{-1/4}
         (S\pa_jS)  \nn\\
 &&\hspace{.6in}+\fr{i}4 (\sqrt{h}h^{ij}-\ve^{ij})\pa_i X^+ H^{-7/4}\fr{H'}{r} \pa_jX^u
        X^m\; (S\g^{um}S)\nn\\
  && \hspace{.6in}   -\fr{i}4 (\sqrt{h}h^{ij}-\ve^{ij})\pa_i X^+ H^{-5/4}\fr{H'}{r}
       \pa_jX^m X^n\;(S\g^{mn}S)
       \left.\fr{}{}\right]\nn\\
       &&+\fr1{4(p^+)^2}\sqrt{h}h^{ij}\pa_i X^+\pa_j
       X^+\;H^{-1/2}\nn\\
      && \left[\fr{}{}\right.
  -\fr{17}{1536}\k_1(S\g^{uv} S)( S\g^{uv} S)+\left(
  \fr{43}{768}\k_1+\fr1{192}\k_2
  \right)(S\g^{au} S)( S\g^{au} S) \nn\\
  &&\hspace{.6in}-\left(
  \fr1{192}\k_2 +\fr1{128}\k_1\right) (S\g^{ab} S)( S\g^{ab} S)\nn\\
  &&\hspace{.6in}+X^aX^b \fr{1}{r^2}\left(\fr{31}{768}\k_1-\fr1{32}\k_2
  \right)(S\g^{au} S)( S\g^{bu} S)\nn\\
  &&\hspace{.6in}+X^aX^b \fr{1}{r^2}\left(
  +\fr1{32}\k_2+\fr{29}{384}\k_1\right)(S\g^{ac} S)( S\g^{bc} S)
  \left.\fr{}{}\right] \label{startingaction5quoted'}
 \eea
where $h^{ij}$ is the world-sheet metric $\eta^{ij}=Diag(-1,1)$ and
 \bea
 H=1+\fr{4\pi g^2\a'^2}{r^4},\quad\quad\k_1=H^{-5/2}(H')^2,\quad\quad \k_2=H^{-3/2}H'\fr1r
 \eea
 In $H$, we have used the open string coupling square, $g^2$, in the place of
  the more commonly used closed string coupling, $g_c$. About the notations, the ten
dimensional bosonic coordinate, $X^M$, consists of
 \bea
 X^M=(\X^\pm, X^u, X^m)
 \eea
where $u$ runs in the two world-volume directions and $m$ or $a$
runs in the six transverse directions with $r^2\equiv \sum
_m(X^m)^2$. (We do not distinguish the index $a$ from the index
$m$.) Below often we use the following abbreviations
 \bea
 A=(u,m)
 \eea
 The fermionic coordinate, $S$, is
 the re-scaled coordinate of the standard fermionic coordinate, $\th$. We refer
 to \cite{Park:2008fp,gsw} for more details on the notations.

 The individual
counter vertex results from the expansion of $e^{-\fr12 TV_G}$ where
$T$ is the string tension. We define\footnote{\ni In later sections,
we also use the notation,
 \bea
 V_G &\equiv& V_{G,q}+V_{G,q^2}+...
 \eea
 In other words, $V_{G,q^n}\equiv V_{G,g^{2n} }$ with $q$ defined in
 (\ref{qdefi}).
  }
 \bea
 V_G &\equiv& V_{G,g^2}+V_{G,g^4}+...\nn\\
 e^{-\fr12TV_G}&=&1-\fr12TV_G+\fr18T^2V_G^2+...
 \eea
For the two-loop order, we use the same strategy: first we compute a
disc with two holes and isolate the divergence from it. The
difference is that one cannot use the light-cone formulation in
order to compute the diagram. However, the tree amplitudes with the
counter vertex operator inserted can still be computed in the
light-cone formulation. As a matter of fact, the light-cone
Green-Schwarz seems the most convenient for that particular purpose.
Fortunately two-loop closed string amplitudes have been computed
recently in other formulations
\cite{Zheng:2002ji}\cite{D'Hoker:2001nj}\cite{Berkovits:2005df}. The
measure for the open string amplitudes can be deduced from the
closed string ones by noting that an open string surface is related
to the corresponding closed string surface by a simple
identification procedure. Including the measure, a two-loop
four-point vector amplitude, for example, in the hyperelliptic
parametrization should take a form of
 \bea
 A(1,2,3,4)=&& K_0\int \fr{da_1da_2da_3}{a_{12}a_{23}a_{31}}
 \fr{1}{\mbox{T}^5} \prod_{i=1}^4 \fr{dz_i}{y(z_i)}
 \prod_{i<j}\exp[-k_i\cdot k_j G(z_i,z_j)]
 \nn\\
 &&\;\;[s(z_1z_2+z_3z_4)+t(z_1z_4+z_2z_3)+u(z_1z_3+z_2z_4)]
 \eea
The result is valid up to an overall numerical constant. $K_0$ is
the usual tree-level kinematic factor defined e.g., in (7.4.42) of
\cite{gsw}. The variables, $a_i$, are the moduli parameters of the
hyperelliptic representation of the two-loop diagram. We refer to
\cite{Zheng:2002ji}) (and the references therein) for the
definitions of $\mbox{T}$ (not to be confused with the string
tension $T$) and other quantities. Based on a survey of literatures
(e.g.,\cite{Zheng:2002ji,D'Hoker:2005jc,DiVecchia:1996uq,Mafra:2008ar})
we expect that the divergence will presumably occur  only when the
locations of all four vertex operators coincide.\footnote{\ni To be
able to see it explicitly it will be necessary to make change of
variables from $z_i$-coordinates to $(q,\n_i)$-coordinates that are
used in e.g., section 8.1 of \cite{gsw}.} It implies then that the
amplitude becomes factorized to contain $s+t+u$ as an overall
factor, which vanishes for massless states: the two-loop amplitude
seems finite unlike the one-loop amplitude. If there were a
divergence we would have tried to cancel it with
$<VVVV\;V_{G,g^4}>_{\mbox{tree}}$\footnote{As a matter of fact, one
should also consider the contribution form $< VVVV
V_{G,g^2}>_{\mbox{1 loop}}$ to be complete as would be expected in
analogy with a normal quantum field theory analysis. This point has
been brought to our attention by a few people. The outcome of such
computation will constitute another crucial test of our conjecture.
Since it is a five-point or more in one-loop order, we will take the
task in some place else in the near future.} where $V_{G,g^4}$ is
the quartic order term in the coupling constant expansion of $V_G$,
which is the order in which the coupling constant appears for the
two loop diagrams. Since there is no divergence to cancel against,
the counter vertex operator should not generate any divergence or at
least any new form of it. If it does not generate any divergence the
"renormalization" procedure will be simpler than otherwise. It turns
out that it does generate divergences but fortunately they come with
the same kinematic factors as the tree amplitudes. As we discuss now
they can be absorbed by re-scaling the external vertex operators.
String tension renormalization may be required as well in general.

At two-loop order, there are two types of the counter-vertex terms.
Schematically they can be written as
 \bea
 (V_{G,g^2})^2\oplus  V_{G,g^4}
 \eea
where $\oplus$ indicates that the precise relative coefficients are
not being recorded. They are obtained by the large-$r_0$ expansion
of Eq.(\ref{startingaction5quoted'}). Various vertices in
$(V_{G,g^2})^2$ and $V_{G,g^4}$ contain terms of the types $\int
d^2z\;(\cdots)$ and $\int d^2z\;(\cdots)\int dz'(\cdots)$. For the
single integral terms, we use a prescription where one freely
performs integration by parts, and only at the final moment one
replaces $\int d^2z \ra \int dy$. For the terms that come with $\int
d^2z\;(\cdots)\int dz'(\cdots)$ one may partially integrate, after
inserting the external vertex operators, to get a delta function.
One of the two $z$-integrations can be removed by the resulting
delta function. (More details are presented in the next section.)
Then we follow the same prescription that is used for the single
integral terms. We show after lengthy algebra in the next section
that it is only the $\pa X\cdot \pa X$-terms (or the terms that
produce the $X\cdot \pa X$-terms after a delta function is used)
that yield non-vanishing results: together with the external vertex
operators, the quartic order counter-vertex operator yields
 \bea
 <VVVV\;[(V_{G,g^2})^2\oplus V_{G,g^4}]>\;\;\propto\;\;
 <VVVV\;\int\pa X\cdot \pa X>
 \eea
More precisely, we show in the next section the following results
for the vector multiplet correlators and the scalar multiplet
correlators \footnote{ The the vector multiplet, $V_v$, and the
scalar multiplet, $V_s$, were obtained in \cite{Park:2007mc} and are
quoted for convenience in one of the appendices.}
 \bea
 <V_v V_v V_v V_v
 \;V_{G,g^4}>&=&-q^2\,\fr{3}{16}<V_v V_v V_vV_v
 \;\int_{x_1}^\infty \; (\pa_i X^u \pa_i X^u) >\nn\\
 <V_v V_v V_v V_v\;(V_{G,g^2})^2>
 &=&\fr{q^2}{8}<V_v V_v V_v V_v\int_{x_1}^\infty\;
     \left(\pa_i X^u \pa_i X^u\right)>\nn\\
 <V_s V_s V_s V_s\;V_{G,g^4}>
 &=&\fr{q^2}{16}<V_s V_s V_s V_s\int_{x_1}^\infty  \;
      (\pa_i X^m \pa_i X^m)>\nn\\
  <V_s V_s V_s V_s\;(V_{G,g^2})^2>
  &=&\fr{q^2}{8}
    <V_s V_s V_s V_s\int_{x_1}^\infty
      \; \left(\pa_i X^m \pa_i X^m\right)>
      \label{correlator-summary}
 \eea
 The definition of the parameter $q$ can be found in \rf{qdefi}.
The explanation of the particular form of the integration can be
found in, e.g., sec3 of \cite{Park:2008fp}. The RHS's yield the tree
diagram kinematic factors with additional divergence factors. These
divergences can be absorbed in the following way. In our convention
the free part of the action is given by
 \bea
 -\fr{T}{2}\int d^2\s\; \pa X\pa X
 \eea
 The free fermionic
 pieces have been omitted: as in the one-loop case \cite{Park:2008fp},
 they only lead to vanishing results. (See also the discussion below \rf{xxp}.)
For the amplitudes that we consider at the given order, it is not
necessary to perform string tension renormalization.
 However, we still consider it just in case we need
  when computing other and/or higher order amplitudes in the near future:
 we write
 \bea
  T&=&T_0+T_{1}g^2+T_{2}g^4+...
 \eea
We also introduce the vertex operator renormalization
 \bea
 V \Rightarrow {\cal N}_V\; V\;\; \mbox{where}\;\;
 {\cal N}_V &=&1+{\cal N}_{V}^\1g^2+{\cal N}_{V}^\2g^4+...
 \eea
 where ${\cal N}_{V} \equiv {\cal N}_{v}$ for the vector operator and
  ${\cal N}_{V} \equiv{\cal N}_{s}$ for the scalar operator.
 The counter correlator is given by
 \bea
 <({\cal N}_V)^4\;VVVV\;e^{-\fr{T-T_0}{2}\int d^2\s\; \pa X \pa X}
         e^{-\fr{T}{2}V_G}>
 \eea
 Writing
 \bea
  V_G&=&V_{G,g^2}+V_{G,g^4}+... \label{VGexpansion}
 \eea
and focusing on the exponential terms one gets
 \bea
 &&({\cal N}_V)^4e^{-\fr{T-T_0}{2}\int d^2\s\; \pa X \pa X}
         e^{-\fr{T}{2}V_G}-1\nn\\
   =&& -\fr12\left(T_{1}g^2\int\pa X\pa X+T_0V_{G,g^2}\right)
     +4{\cal N}_{V}^\1\;g^2\nn\\
 &&  -\fr12\left(T_{2}g^4\int\pa X\pa X
           +T_{1}g^2V_{G,g^2}+T_0V_{G,g^4}\right)
            +\fr18\left(T_{1}g^2\int\pa X\pa X+T_0V_{G,g^2}\right)^2
            +...\nn\\
   && +4{\cal N}_{V}^\1\;g^2\;
         \left(-\fr12\right)\left(T_{1}g^2\int\pa X\pa X+T_0V_{G,g^2}\right)
   +(6[{\cal N}_{V}^\1]^2+4{\cal N}_{V}^\2)g^4
            \label{vertices}
 \eea
 The first line should cancel the one-loop divergence. It
 implies that
 \bea
 \fr{g^2}{4}\fr{1}{\e'} +\fr12\fr{1}{\e}\left(T_{1}g^2-\fr{q}{2}T_0\right)
     +4{\cal N}_{V}^\1\;g^2=0 \label{one-loop}
 \eea
 where the first term is the one-loop divergence. The symbol, $\e'$, is
 the cutoff that was used in the one-loop computation
 \cite{Park:2008fp}. (It is called $\e_y$ in Eq.(25) of
 \cite{Park:2008fp}.) Similarly $\e$ is associated with the two-loop
 terms.\footnote{As with our previous work \cite{Park:2008fp}, we use dimensional
 regularization and world-sheet integration cutoff together. It seems that it is not unusual
 in string theory to use two different regularization methods at the same time. For
 example, a combined use of normal ordering and string tension renormalization is made in
\cite{gsw}. Although it is not certain at this point, it might be
that dimensional regularization should be viewed as a world-sheet
regularization whereas the cut-off together with the deformed
geometry should be viewed as a space-time renormalization.} The fact
that we have assigned a different regulator, $\e$, for the two loop
is matter of convenience: it is to keep track of the one loop and
the two loop computations separately. At any point, $\e'$ can be set
$\e'=\e$.
 The loop expansion parameter
  $q$ is defined as
 \bea
 q&=&\fr{4\pi g^2  \a'^2}{r_0^4} \label{qdefiquote}
 \eea
 in (\ref{qdefi}), below which the definition of $r_0$ can be found
 as well. We also have used
 \bea
 <VVVV\;\pa X\pa X>&=&-\fr{1}{\e}\;<VVVV>
 \eea
Since eq(\ref{one-loop}) is valid both for the vector and the
multiplets, it implies
 \bea
 {\cal N}_{v}^\1={\cal N}_{s}^\1 \equiv {\cal N}^\1
 \label{rnconst-rel1}
 \eea
After substituting (\ref{qdefiquote}) into (\ref{one-loop}) one gets
 \bea
  {\cal N}^\1&=&-\fr{1}{16\,\e'}
            -\fr{1}{8\,\e}\;T_{1}+\fr{1}{16\,\e}\;\fr{\pi}{r_0^4}T_0
 \label{rnconst-rel2}
  \eea
 The second and third lines of (\ref{vertices}) together should remove
 the two loop divergence. As we have discussed above, the two-loop
 amplitudes seem finite. Therefore we require the correlator that involves
 the second and the third line to vanish when computed with the four
external vertex operators inserted. For the vector states it yields
 \bea
 &&<\left\{\fr{}{}\right.\fr12\left(T_{2}g^4\int\pa X\pa X
           +T_{1}g^2V_{G,g^2}+T_0V_{G,g^4}\right) \nn\\
  && -\fr18\left(T_{1}^2g^4(\int\pa X\pa X)^2
             +2T_{1}g^2(\int\pa X\pa X)T_0V_{G,g^2}
             +T_0^2(V_{G,g^2})^2\right)
    \nn\\
    && - \left[-2{\cal N}_{v}^\1\;g^2\;
         \left(T_{1}g^2\int\pa X\pa X+T_0V_{G,g^2}\right)
   +(6[{\cal N}_{v}^\1]^2+4{\cal N}_{v}^\2)g^4\right]\left.\fr{}{}\right\}V_v V_v V_v
      V_v>\nn\\
   &&=0
 \eea
 One gets after using (\ref{correlator-summary}) and the results
 of \cite{Park:2008sg,Park:2008fp}
 \bea
 &&\k\left[g^4T_{2}-\fr{q}{2}g^2T_{1}-q^2\fr{3}{16}T_0
  -\fr14\left(-2cg^4T_{1}^2
             +2cqg^2T_{1}T_0
             -c\fr{q^2}{8}T_0^2\right)\right.\nn\\
 &&\left.  \quad\quad +4{\cal N}_{v}^\1\;g^2\;
         \left(T_{1}g^2-\fr{q}{2}T_0\right)\right]
         +2(6[{\cal N}_{v}^\1]^2+4{\cal N}_{v}^\2)g^4
            \label{2loop-v}
 =0
 \eea
 Substituting (\ref{qdefiquote}) and dividing by $g^4$ it simplifies
 to
 \bea
 &&-\fr{1}{\e}\left[T_{2}-\fr{\pi}{2r_0^4}T_{1}-\fr{3\pi^2}{16r_0^8}T_0
  -\fr12T_{1}^2
             +\fr12\fr{\pi}{r_0^4}T_{1}T_0
             -\fr{\pi^2}{32r_0^8}T_0^2\right.\nn\\
 &&\left.  \quad\quad +4{\cal N}_{v}^\1\;
         \left(T_{1}-\fr{\pi}{2r_0^4}T_0\right)\right]
         +12({\cal N}_{v}^\1)^2+8{\cal N}_{v}^\2
 =0 \label{rnconst-rel3}
 \eea
Similarly, requiring
 \bea
 &&<\left\{\fr{}{}\right.\fr12\left(T_{2}g^4\int\pa X\pa X
           +T_{1}g^2V_{G,g^2}+T_0V_{G,g^4}\right) \nn\\
  && -\fr18\left(T_{1}^2g^4(\int\pa X\pa X)^2
             +2T_{1}g^2(\int\pa X\pa X)T_0V_{G,g^2}
             +T_0^2(V_{G,g^2})^2\right)
    \nn\\
    && - \left[-2{\cal N}_{v}^\1\;g^2\;
         \left(T_{1}g^2\int\pa X\pa X+T_0V_{G,g^2}\right)
   +(6[{\cal N}_{v}^\1]^2+4{\cal N}_{v}^\2)g^4\right]\left.\fr{}{}\right\}
   V_s V_s V_s V_s>\nn\\
   &&=0
 \eea
for the scalar states leads to
 \bea
 &&-\fr{1}{\e}\left[T_{2}-\fr{\pi}{2r_0^4}T_{1}+\fr{\pi^2}{16r_0^8}T_0
  -\fr12T_{1}^2
             +\fr12\fr{\pi}{r_0^4}T_{1}T_0
             -\fr{\pi^2}{32r_0^8}T_0^2\right.\nn\\
 &&\left.  \quad\quad +4{\cal N}_{s}^\1\;
         \left(T_{1}-\fr{\pi}{2r_0^4}T_0\right)\right]
         +12({\cal N}_{s}^\1)^2+8{\cal N}_{s}^\2
 =0 \label{rnconst-rel4}
 \eea
By subtracting (\ref{rnconst-rel4}) from (\ref{rnconst-rel3}) one
gets
 \bea
-{\cal N}_{s}^\2+{\cal
N}_{v}^\2-\fr{1}{32\,\e}\fr{\pi^2}{r_0^8}T_0=0 \label{rnconst-rel5}
 \eea
Therefore it is verified that the one- and two- loop divergences can
be absorbed by the operator renormalization relations
(\ref{rnconst-rel1}), (\ref{rnconst-rel2}), (\ref{rnconst-rel3}),
(\ref{rnconst-rel4}) and (\ref{rnconst-rel5}).

\section{Two-loop counter terms}

 In this section, we prove the results given
 in (\ref{correlator-summary}). The geometry vertex operator, $V_G$, given in
(\ref{startingaction5quoted'}) is the unique counter vertex operator
in the sense that the form does not depend on the external
scattering states and the number of loops. This is true at least for
the correlators that we have computed so far, which is rather
remarkable. Below we expand $V_G$ in terms of two parameters,
$g,r_0$. The parameter $q$ is defined as
 \bea
 && q=\fr{4\pi g^2  \a'^2}{r_0^4} \label{qdefi}
 \eea
where $r_0^2=\sum_m (X_0^m)^2$ with $X_0$ appearing in the shift,
$X^m\ra X^m+X_0^m$ \cite{Park:2008fp}. It counts the number of loops
among other things. Since we are interested in the two-loop
amplitudes, we select the terms with $q^2$. As for the $r_0$, we
keep the terms up to (and including) $\fr{1}{r_0^{10}}$ order. With
the expanded vertex operator inserted we consider in the next
section two scattering amplitudes: the four vector scattering
amplitude and the four scalar scattering amplitude. As emphasized in
the previous works (e.g., \cite{Park:2008fp}), the computations are
at the tree level.

\ni There are two terms that appear in (\ref{VGexpansion}) at this
order: the q-order vertex operator, $V_{G,q}$, and the $q^2$-order
vertex operator, $V_{G,q^2}$. One can show by straightforward
algebra that they are respectively given by\footnote{As mentioned in
one of the previous footnotes, our notation is such that
$V_{G,q^n}\equiv V_{G,g^{2n} }$.}
 \bea
  V_{G,q}
 =&&{q}\int  \;(-)\fr{1}2\sqrt{h}\;h^{ij}
  \left[\pa_i X^u \pa_j X^v \eta_{uv}
     \left(  - \frac{1}{2}
  + \frac{2\,X_0\cdot X}{{r_0}^2} + \frac{r^2}{{r_0}^2}
  - \frac{6\,{\left( X_0\cdot X \right) }^2}{{r_0}^4}
     \right) \right.\nn\\
   &&\left.\hspace{1.2in}
     +\pa_i X^m \pa_j X^n \eta_{mn}
     \left(  \frac{1}{2}-\frac{2 X_0.X }{{r_0}^2}
 +\frac{6(X_0.X)^2}{{r_0}^4}-\frac{r^2}{{r_0}^2}\right)\right]
  \nn\\
 && +\fr1{2p^+}(\sqrt{h}\;h^{ij}-\ve^{ij})\pa_i X^+
               \left[\fr{}{}\right.
   -2i(S\pa_jS)\left(-\frac{1}{4}+\frac{X_0.X }{r_0^2}
            +\frac{r^2}{2 r_0^2}-\frac{3
   (X_0.X)^2}{r_0^4}\right)\nn\\
 &&-{i}  \pa_jX^u \fr{(X+X_0)^m}{r_0^2}\; (S\g^{um}S)
       +{i} \pa_jX^m \fr{(X+X_0)^n}{r_0^2}\;(S\g^{mn}S)
      \left.\fr{}{} \right] \nn\\
       &&+\fr1{4(p^+)^2}\sqrt{h}h^{ij}\pa_i X^+\pa_j X^+\;
      \frac{1}{r_0^2}  \left[\fr{}{}\right.
  -\fr1{48}(S\g^{au} S)( S\g^{au} S)
  +\fr1{48}(S\g^{ab} S)( S\g^{ab} S)\nn\\
 &&  + \fr{(X+X_0)^a(X+X_0)^b}{8r_0^2}
      \left((S\g^{au} S)( S\g^{bu} S)
  -\fr1{8}(S\g^{ac} S)( S\g^{bc} S)\right)
 \left.\fr{}{}\right]
 \label{vqmother}
 \eea
 \bea
 V_{G,q^2}
 =&&{q^2}\int  \;(-)\fr{1}2\sqrt{h}\;h^{ij}
      \left[\pa_i X^u \pa_j X^v \eta_{uv}
     \left(  \frac{3}{8}
   - \frac{3\,X_0\cdot X}{{r_0}^2}-  \frac{3\,r^2}{2\,{r_0}^2}
       + \frac{15\,{\left( X_0\cdot X \right) }^2}{{r_0}^4}
     \right) \right.\nn\\
   &&\left. +\pa_i X^m \pa_j X^n \eta_{mn}
     \left(   -\frac{1}{8}+\frac{X_0.X }{{r_0}^2}
  +\frac{r^2}{2 {r_0}^2}
   -\frac{5(X_0.X)^2}{{r_0}^4}\right) \right]
  \nn\\
 && +\fr1{2p^+}(\sqrt{h}\;h^{ij}-\ve^{ij})\pa_i X^+
     \left[\fr{}{}\right.
  -2i (S\pa_jS)\left(\frac{5}{32}-\frac{5 X_0.X }{4 r_0^2}
           + \frac{25 (X_0.X)^2}{4 r_0^4}
   -\frac{5 r^2}{8 r_0^2} \right) \nn\\
 &&+\fr{7i}4\;  \pa_jX^u\fr{(X+X_0)^m}{r_0^2}\; (S\g^{um}S)
     -\fr{5i}4\; \pa_jX^m \fr{(X+X_0)^n}{r_0^2}\;(S\g^{mn}S)
        \left.\fr{}{}\right] \nn\\
       &&+\fr1{4(p^+)^2}\sqrt{h}h^{ij}\pa_i X^+\pa_j X^+
           \frac{1}{r_0^2} \;\left[\fr{}{}\right.
  -\fr{17}{96}(S\g^{uv} S)( S\g^{uv} S)
   +\fr{45}{48}(S\g^{au} S)( S\g^{au} S)\nn\\
  &&\hspace{2.5in}-\fr1{6}(S\g^{ab} S)( S\g^{ab} S)\nn\\
  && + \fr{(X+X_0)^a(X+X_0)^b}{r_0^2}\left(
     \fr{19}{48}(S\g^{au} S)( S\g^{bu} S)
    +\fr{35}{24}(S\g^{ac} S)( S\g^{bc} S)\right)
    \left.\fr{}{}\right]           \label{vqqmother}
  \eea
The task now is to compute the four correlators: $<V_v V_v V_v V_v
 \;V_{G,g^4}>,\;\;<V_v V_v V_v V_v\;(V_{G,g^2})^2>,\;\;<V_s V_s V_s V_s\;V_{G,g^4}>$
 and $<V_s V_s V_s V_s\;(V_{G,g^2})^2>$. There are some terms in (\ref{vqmother}) and
 (\ref{vqqmother}) that do not contribute to any of these
 correlators for obvious reasons. Below we will explain them with
 specific examples.
In the actual computations with the individual vertices in
(\ref{vqmother}) and (\ref{vqqmother}), many of them turn out to
vanish for certain reasons. Some of the contractions need not be
executed as dimensional regularization is being used. Some
contractions produce delta functions due to the relation,
 \bea
 <\pa ^2 X(x)X(x')>\sim \d(x-x')
 \eea
If both of the arguments are, for example, those of the external
states, the contraction will vanish since the locations of the
external vertex operators are such that $x_1>1>x>0$. The only
potentially non-zero contractions come from the following sets,
 \bea
 (x,x') =(x_1,y)\;\;,\;\; (y,x_1)\;\;, \;\; (y,y')\label{xxp}
 \eea
 where the $y$'s are the arguments for the geometry vertices.
A delta function also appears when $(\sqrt{h}\;h^{ij}-\ve^{ij})\pa_j
S$ contracts with another $S$.
 It is why, as in the one-loop case, the $S\pa_j S$-terms in
 (\ref{vqmother}) and (\ref{vqqmother}) may be dropped:
to avoid a vanishing delta both factors of $S$ need to contract with
the fermionic part of the external vertex operator,
$S(x_1)\g^{ij}S(x_1)$. It yields $\Tr \g^{A_1B_1}=0$: henceforth the
$S\pa S$-term will be omitted. Some of the terms appear with
$\fr{1}{x_1^3}$ or higher power. They vanish since the measure
contains only $x_1^2$ and it is taken $x_1\ra \infty$ at the end.
There are other terms that have an overall factor of either $k_1^2$,
$k_1\cdot \z_1$, or $s+t+u$. They vanish due to the masslessness of
the external states, the polarization constraint, or momentum
conservation. Below we examine each of the terms in (\ref{vqmother})
and (\ref{vqqmother}) together with the four external vertex
operators. As just described, the majority of terms
vanish due to one or more of the following reasons:\\

\ni $\bullet$ dimensional regularization\\
 $\bullet$ vanishing $\d$-function\\
 $\bullet$ index structures\\
 $\bullet$ sub-leading in $\fr{1}{x_1}$\\
 $\bullet$ momentum conservation\\
 $\bullet$ overall $k_1^2$ or $k_1\cdot \z_1$ factor\\

\subsection{Four vector scattering}

For convenience, we record the explicit form of the product of four
vector vertex operators,
 \bea
&& V^{u_1}_v(x_1) V^{u_2}_v(x_2) V^{u_3}_v(x_3) V^{u_4}_v(x_4) \nn\\
&& =\dot{X}^{u_1}\dot{X}^{u_2}\dot{X}^{u_3}\dot{X}^{u_4}
+l^8R^{u_1v_1}k_1^{v_1}R^{u_2v_2}k_2^{v_2}R^{u_3v_3}k_3^{v_3}R^{u_4v_4}k_4^{v_4}\nn\\
&&-l^2\left[\dot{X}^{u_1}\dot{X}^{u_2}\dot{X}^{u_3}R^{u_4v_4}k_4^{v_4}
  +\dot{X}^{u_1}\dot{X}^{u_2}\dot{X}^{u_4}R^{u_3v_3}k_3^{v_3}\right.\nn\\
&&\left.\;\;\;\;+\dot{X}^{u_1}\dot{X}^{u_3}\dot{X}^{u_4}R^{u_2v_2}k_2^{v_2}
   +\dot{X}^{u_2}\dot{X}^{u_3}\dot{X}^{u_4}R^{u_1v_1}k_1^{v_1}\right]\nn\\
&&+l^4\left[\dot{X}^{u_1}\dot{X}^{u_2}R^{u_3v_3}k_3^{v_3}R^{u_4v_4}k_4^{v_4}
  +\dot{X}^{u_3}\dot{X}^{u_4}R^{u_1v_1}k_1^{v_1}R^{u_2v_2}k_2^{v_2}\right.\nn\\
 && \left.\;\;\;\;+\dot{X}^{u_1}\dot{X}^{u_4}R^{u_2v_2}k_2^{v_2}R^{u_3v_3}k_3^{v_3}
  +\dot{X}^{u_1}\dot{X}^{u_3}R^{u_2v_2}k_2^{v_2}R^{u_4v_4}k_4^{v_4}\right.\nn\\
  &&\left. \;\;\;\;+\dot{X}^{u_2}\dot{X}^{u_3}R^{u_1v_1}k_1^{v_1}R^{u_4v_4}k_4^{v_4}
  +\dot{X}^{u_2}\dot{X}^{u_4}R^{u_1v_1}k_1^{v_1}R^{u_3v_3}k_3^{v_3}\right]
  \nn\\
 &&-l^6\left[\dot{X}^{u_1}R^{u_2v_2}k_2^{v_2}R^{u_3v_3}k_3^{v_3}R^{u_4v_4}k_4^{v_4}
        +\dot{X}^{u_2}R^{u_1v_1}k_1^{v_1}R^{u_3v_3}k_3^{v_3}R^{u_4v_4}k_4^{v_4}
        \right. \nn\\
&&\left.\;\;\;\;+\dot{X}^{u_3}R^{u_1v_1}k_1^{v_1}R^{u_2v_2}k_2^{v_2}R^{u_4v_4}k_4^{v_4}
        +\dot{X}^{u_4}R^{u_1v_1}k_1^{v_1}R^{u_2v_2}k_2^{v_2}R^{u_3v_3}k_3^{v_3}
        \right]\label{4v}
 \eea
We loosely refer to each type of these terms as XXXX, RRRR, XXXR,
XXRR, and XRRR respectively. The parameter $l$ is the standard one
given by $l=\sqrt{2\a'}$. In some places we set $l=1$.

\subsubsection{ $V_{G,q^2}$ contribution }

For the four vector scattering the vertex $V_{G,q^2}$ given in
(\ref{vqqmother}) simplifies to
 \bea
   V_{G,q^2}
 =&&q^2\int  \;-\;\fr{3}{16}\sqrt{h}\;h^{ij}
      \pa_i X^u \pa_j X^v \eta_{uv}
  - \frac{5}{32}\fr{i}{p^+}
   (\sqrt{h}\;h^{ij}-\ve^{ij})\pa_i X^+(S\pa_jS)
         \nn\\
       &&+\fr1{4(p^+)^2r_0^2}\sqrt{h}h^{ij}\pa_i X^+\pa_j X^+\;
       \left[\fr{}{}\right.\nn\\
  && -\fr{17}{96}(S\g^{uv} S)( S\g^{uv} S)
  +\fr{15}{16}(S\g^{au} S)( S\g^{au} S)
  -\fr16(S\g^{ab} S)( S\g^{ab} S)\nn\\
  &&
 + \fr{19}{48}\fr{X_0^aX_0^b}{r_0^2}
  (S\g^{au} S)( S\g^{bu} S)
  + \fr{35}{24}\fr{X_0^aX_0^b}{r_0^2}
 (S\g^{ac} S)( S\g^{bc} S)
 \left. \fr{}{}\right]
 \eea
 This is due to the fact that all the terms that contain $X^m$
 drop due to dimensional regularization. Also
 some of the terms were shown to vanish in the one-loop analysis
 \cite{Park:2008fp}. Consider\footnote{We refer to \cite{Park:2008fp}
 for simpler sample calculations.}
 \bea
 &&<V^{u_1}_v(x_1) V^{u_2}_v(x_2) V^{u_3}_v(x_3) V^{u_4}_v(x_4)
 ;V_{G,q^2}>
 \eea
 The $\pa X \pa X$-term gives the tree level kinematic
factor. We now show that none of the other terms makes a
contribution. Let's focus on the terms with four $S$'s:
 \bea
 &&<V^{u_1}_v(x_1) V^{u_2}_v(x_2) V^{u_3}_v(x_3) V^{u_4}_v(x_4)
 ;S\g^{AB}SS\g^{CD}S>\nn\\
 =&&< \left[\fr{}{}\right.
l^4\left(\dot{X}^{u_1}\dot{X}^{u_2}R^{u_3v_3}k_3^{v_3}R^{u_4v_4}k_4^{v_4}
  +\dot{X}^{u_3}\dot{X}^{u_4}R^{u_1v_1}k_1^{v_1}R^{u_2v_2}k_2^{v_2}\right.\nn\\
 && \left.\;\;\;\;+\dot{X}^{u_1}\dot{X}^{u_4}R^{u_2v_2}k_2^{v_2}R^{u_3v_3}k_3^{v_3}
  +\dot{X}^{u_1}\dot{X}^{u_3}R^{u_2v_2}k_2^{v_2}R^{u_4v_4}k_4^{v_4}\right.\nn\\
  &&\left. \;\;\;\;+\dot{X}^{u_2}\dot{X}^{u_3}R^{u_1v_1}k_1^{v_1}R^{u_4v_4}k_4^{v_4}
  +\dot{X}^{u_2}\dot{X}^{u_4}R^{u_1v_1}k_1^{v_1}R^{u_3v_3}k_3^{v_3}\right)
  \nn\\
 &&-l^6\left(\dot{X}^{u_1}R^{u_2v_2}k_2^{v_2}R^{u_3v_3}k_3^{v_3}R^{u_4v_4}k_4^{v_4}
        +\dot{X}^{u_2}R^{u_1v_1}k_1^{v_1}R^{u_3v_3}k_3^{v_3}R^{u_4v_4}k_4^{v_4}
        \right. \nn\\
&&\left.\;\;\;\;+\dot{X}^{u_3}R^{u_1v_1}k_1^{v_1}R^{u_2v_2}k_2^{v_2}R^{u_4v_4}k_4^{v_4}
        +\dot{X}^{u_4}R^{u_1v_1}k_1^{v_1}R^{u_2v_2}k_2^{v_2}R^{u_3v_3}k_3^{v_3}
        \right)\nn\\
&& +l^8R^{u_1v_1}k_1^{v_1}R^{u_2v_2}k_2^{v_2}
R^{u_3v_3}k_3^{v_3}R^{u_4v_4}k_4^{v_4} \left.\fr{}{}\right]
  \nn\\
  &&\quad\quad\;S\g^{AB}S\;S\g^{CD}S>\label{4szero}
 \eea
One can show that these $SSSS$-terms yield vanishing expressions.
For example, by explicit calculation one can verify that
 \bea
 &&<V^{u_1}_v(x_1) V^{u_2}_v(x_2) V^{u_3}_v(x_3) V^{u_4}_v(x_4)
 ;S\g^{au}SS\g^{bu}S>=0
 \eea
All the other remaining terms turn out to give vanishing
expressions. For an illustration, let us take the terms that contain
a factor, $(\z_2\cdot \z_3)$. One can show after the correlators  in
(\ref{4szero}) are explicitly calculated that the coefficient of
$(\z_2\cdot \z_3)$ is proportional to
 \bea
   &&  \left\{ \left( \fr{q^2}{r_0^2}\right)
  \left[-\fr{17}{96}(S\g^{uv} S)( S\g^{uv} S)
  -\fr16(S\g^{ab} S)( S\g^{ab} S)\right]
   +\left(\fr{q^2}{r_0^4}\right) X_0^aX_0^b
  \left[\fr{35}{24}(S\g^{ac} S)( S\g^{bc} S)
  \right]\right\}\nn\\
\Rightarrow &&
 -2 \frac{  1}{(1-x)^2}
   \z_1\cdot k_4\;\z_4\cdot k_1
  +\left( \z_4\cdot k_2
       +\frac{ \z_4\cdot k_3}{x}\right)
       \frac{1 } { 1-x }
        (\z_1\cdot k_2u-\z_1\cdot k_3s)
       \nn\\
  && -
 \frac{1}{{\left( 1 - x \right) }^2\,}
      \;t\z_1\cdot k_4\;\z_4\cdot k_1
     \nn\\
  && -\frac{1}{2x  }
     \left(s\z_1\cdot k_2\; \z_4\cdot k_1
       -t\z_1\cdot k_2\; \z_4\cdot k_3
     +s\z_1\cdot k_4\; \z_4\cdot k_3
      -t\z_1\cdot k_4\; \z_4\cdot k_1 \right.\nn\\
 &&\left.\hspace{.4in}+u\z_1\cdot k_4\; \z_4\cdot k_2
  +u\z_1\cdot k_3\; \z_4\cdot k_1
   -t\z_1\cdot k_3\; \z_4\cdot k_2 \right)\nn\\
  && +\frac{1}{2x\left( 1 -x \right) \, }
       (-s\z_1\cdot k_2\; \z_4\cdot k_1
        +t\z_1\cdot k_2\; \z_4\cdot k_3
     +u\z_1\cdot k_3\; \z_4\cdot k_1 \nn\\
 &&\hspace{.4in}-t\z_1\cdot k_3\; \z_4\cdot k_2
   -s\z_1\cdot k_4\; \z_4\cdot k_3
  +u\z_1\cdot k_4\; \z_4\cdot k_2
   -t\z_1\cdot k_4\; \z_4\cdot k_1 ) \nn\\
  &&- \frac{1}{2\left( 1 -x \right)  }
       (-u\z_1\cdot k_3\; \z_4\cdot k_1
       +t\z_1\cdot k_3\; \z_4\cdot k_2
     +s\z_1\cdot k_2\; \z_4\cdot k_1 -t\z_1\cdot k_2\; \z_4\cdot k_3\nn\\
 &&\hspace{.8in}
           -u\z_1\cdot k_4\; \z_4\cdot k_2
  +s\z_1\cdot k_4\; \z_4\cdot k_3
   -t\z_1\cdot k_4\; \z_4\cdot k_1 )
 \eea
where the Wick rotation has been taken into account. It vanishes
after the $x$-integration. All the other type of terms such as
$(\z_2\cdot \z_4)$ etc and $(\z\cdot k)(\z\cdot k)(\z\cdot
k)(\z\cdot k)$ vanish. Additional examples are presented in one of
the appendices. Therefore one gets
 \bea
V_{G,q^2} \Rightarrow
 \;\;(-q^2)\int  \;\fr{3}{16}\sqrt{h}\;h^{ij}
      \pa_i X^u \pa_j X^v \eta_{uv}
      \label{vec-vvvv-vqq}
 \eea

\subsubsection{$(V_{G,q})^2$ contribution }

\indent Here again, the terms with the factor, $S\pa S$, drop as one
can see as follows. The fermionic parts of the correlators that we
need to consider contain the contraction of the type
 \bea
 S(y)\pa S(y)\; S(y')\pa S(y')\;S(x_i)\g^{u_iv_i}S(x_i)
 \eea
 or
 \bea
 S(y)\pa S(y)\; S(y')\g^{AB} S(y')\;S(x_i)\g^{u_iv_i}S(x_i)
 \eea
 Let's consider the first one. When $\pa S(y)$ and $S(y')$ contract
 the remaining contractions yield a vanishing result. Similarly
 the contraction between $\pa S(y)$ and $S(x_i)$ leads to a null result.
Consider the second equation. Eventually it is necessary to have
$(A,B)=(u,v)$ for a non-vanishing trace of product of $\g$-matrices.
However, such a term is absent in $V_{G,q}$.

  The term with $ \pa_jX^u X^m\; (S\g^{um}S) $ vanishes since it
can only be multiplied with $-\pa_i X^u \pa_j X^v \eta_{uv}+\pa_i
X^m \pa_j X^n \eta_{mn} $, and the number of $X^m$'s is odd. For the
$ \pa_jX^u X_0^m\; (S\g^{um}S) $-term, it is necessary to consider
XXXR-, XXRR-, XRRR- and RRRR-terms when it gets multiplied with
$-\pa_i X^u \pa_j X^v \eta_{uv}+\pa_i X^m \pa_j X^n \eta_{mn} $. The
index structures remove all of them. When it gets squared, i.e., $
\pa_jX^u X_0^m\; (S\g^{um}S)\;\pa_j'X^{u'} X_0^{m'}\; (S\g^{u'm'}S)
$, it seems more convenient to partially integrate to put it into $
X^u X_0^m\; (\pa_jS\g^{um}S)\;X^{u'} X_0^{m'}\; (\pa_j'S\g^{u'm'}S)
$. The fermionic part of the correlators will be of the form
 \bea
 (\pa_jS\g^{um}S) (\pa_j'S\g^{u'm'}S)\; S\g^{u_1v_1}S
 \eea
When $\pa_jS$ contracts with the next term, the remaining
contractions vanish due to dimensional regularization. When it
contracts with the third factor, it is equivalent to the previous
case unless $(\pa_jS\g^{um}S)$ entirely contracts with
$S\g^{u_1v_1}S$, which yields zero.

Consider the terms with $ \pa_jX^m (X+X_0)^n\;(S\g^{mn}S)$. Within
the order that we are considering, $\pa_jX^m X^n$ can only be
multiplied with $\pa_i X^m \pa_j X^n \eta_{mn}$. The correlator
vanishes due to the index structure:
 $m$ and $n$ must be the same while $m'$ and $n'$ must
 be different.
Consider $\pa_jX^m X_0^n(S\g^{mn}S)$-term. The only cross
 term that is relevant is
 \bea
 \pa_jX^{m'} X_0^{n'}(S\g^{m'n'}S)\pa_i X^u \pa_j X^v \eta_{uv}\;
        \frac{2\,X_0\cdot X}{{r_0}^2}
 \eea
Potentially non-vanishing terms in (\ref{4v}) are of the types,
XXXR, XXRR, XRRR, and RRRR. Fermionic index structures kill them
all. Consider the diagonal term
 \bea
 \pa_jX^{m'} X_0^{n'}(S\g^{m'n'}S)\;\pa_jX^m X_0^n(S\g^{mn}S)
 \eea
 By partial integration one gets e.g., $\pa ((S\g^{m'n'}S))$. Due to
 the fermionic index structures, $\pa ((S\g^{m'n'}S))$ can only
 contract with $(S\g^{mn}S)$ but not with external states. It will
 produce a delta function. Then the X-correlator vanishes due to dimensional
 regularization upon using the delta function.

 With these results, the vertex operator can be tailored to
 \bea
 V_{G,q}
 =&& {q}\int  \;(-)\fr{1}2\sqrt{h}\;h^{ij}
  \nn\\
 &&\left(\pa_i X^u \pa_j X^v \eta_{uv}
     \left[  - \frac{1}{2}
  + \frac{2\,X_0\cdot X}{{r_0}^2} +
     \frac{r^2}{{r_0}^2}
  - \frac{6\,{\left( X_0\cdot X \right) }^2}{{r_0}^4}
     \right] \right.\nn\\
   &&\left.
     +\pa_i X^m \pa_j X^n \eta_{mn}
     \left[  \frac{1}{2}-\frac{2 X_0.X }{{r_0}^2}
 +\frac{6(X_0.X)^2}{{r_0}^4}-\frac{r^2}{{r_0}^2}
  \right]\right)\nn\\
       &&-\fr1{(p^+)^2}\sqrt{h}h^{ij}\pa_i X^+\pa_j X^+
   \frac{1}{r_0^2}
       \;\left[\fr{}{}\right.\nn\\
  &&\quad+\fr{1}{192}(S\g^{au} S)( S\g^{au} S)
  -\fr{1}{192}(S\g^{ab} S)( S\g^{ab} S)
   \nn\\
 &&\quad  -\fr1{32}(S\g^{au} S)( S\g^{bu} S)
     \fr{X_0^aX_0^b}{r_0^2}
  +\fr1{32}(S\g^{ac} S)( S\g^{bc} S)
     \fr{X_0^aX_0^b}{r_0^2}
   \left.\fr{}{}\right] \nn\\
  =&&{q}\int  \;(-)\fr{1}2\sqrt{h}\;h^{ij}
  \left[\fr{}{}\right. -\fr12\left(\pa_i X^u \pa_j X^v \eta_{uv}
      -\pa_i X^m \pa_j X^n \eta_{mn} \right)  \nn\\
 && +  \frac{2\,X_0\cdot X}{{r_0}^2}
     \left(\pa_i X^u \pa_j X^v \eta_{uv}
      -\pa_i X^m \pa_j X^n \eta_{mn}\right)
  \nn\\
 &&+
     \left( \frac{r^2}{{r_0}^2}
  - \frac{6\,{\left( X_0\cdot X \right) }^2}{{r_0}^4}
    \right)
     \left(  \pa_i X^u \pa_j X^v \eta_{uv}
            -\pa_i X^m \pa_j X^n \eta_{mn}\right)
  \left.\fr{}{}\right]
    \nn\\
       &&-\fr1{(p^+)^2}\sqrt{h}h^{ij}\pa_i X^+\pa_j X^+
      \fr{1}{r_0^2} \;\left(\fr{}{}\right.\nn\\
  &&\quad+\fr{1}{192}(S\g^{au} S)( S\g^{au} S)
  -\fr{1}{192}(S\g^{ab} S)( S\g^{ab} S)
   \nn\\
 &&\quad  -\fr1{32}(S\g^{au} S)( S\g^{bu} S)
     \fr{X_0^aX_0^b}{r_0^2}
  +\fr1{32}(S\g^{ac} S)( S\g^{bc} S)
  \fr{X_0^aX_0^b}{r_0^2}
   \left.\fr{}{}\right) \nn\\
 \eea
 where in the second equality we have collected the same order terms
 in $\fr{1}{r_0}$. Squaring this, one gets, upon dropping some terms
 that contain the odd number of $X^m$'s,
 \bea
  (V_{G,q})^2
 =&&{q^2}\int \int \left[\fr{}{}\right. \fr14\;\sqrt{h}\;h^{i'j'}
                  \;\sqrt{h}\;h^{ij}  \nn\\
  && +\fr14  \left(\pa_{i'} X^{u'} \pa_{j'} X^{u'}
      -\pa_{i'} X^{m'} \pa_{j'} X^{m'}  \right)
     \left(\pa_i X^u \pa_j X^u
      -\pa_i X^m \pa_j X^m  \right)\nn\\
   &&+4\frac{\,X_0^{l'} X^{l'}}{{r_0}^2}
     \left(\pa_{i'} X^{u'} \pa_{j'} X^{u'}
      -\pa_{i'} X^{m'} \pa_{j'} X^{m'} \right)\nn\\
    &&\quad\;\;  \frac{\,X_0\cdot X}{{r_0}^2}
     \left(\pa_i X^u \pa_j X^u
      -\pa_i X^m \pa_j X^m \right)\left.\fr{}{}\right]\nn\\
 && -\fr12\sqrt{h}\;h^{i'j'}\left(\pa_{i'} X^{u'} \pa_{j'} X^{u'}
      -\pa_{i'} X^{m'} \pa_{j'} X^{m'}  \right)  \nn\\
 &&\left[\fr{}{}\right.
     \fr12\sqrt{h}\;h^{ij}\left( \frac{r^2}{{r_0}^2}
  - \frac{6\,{\left( X_0\cdot X \right) }^2}{{r_0}^4}
    \right)
     \left(  \pa_i X^u \pa_j X^u
            -\pa_i X^m \pa_j X^m \right)
    \nn\\
       &&-\fr1{(p^+)^2}\sqrt{h}h^{ij}\pa_i X^+\pa_j X^+
       \fr{1}{r_0^2}  \;\left(\fr{}{}\right.\fr{1}{192}(S\g^{au} S)( S\g^{au} S)
  -\fr{1}{192}(S\g^{ab} S)( S\g^{ab} S)
   \nn\\
 &&\quad  -\fr1{32}(S\g^{au} S)( S\g^{bu} S)
     \fr{X_0^aX_0^b}{r_0^2}
  +\fr1{32}(S\g^{ac} S)( S\g^{bc} S)
  \fr{X_0^aX_0^b}{r_0^2}
   \left.\fr{}{}\right)
   \left.\fr{}{}\right] \label{VqVq-vec}
 \eea
Let's focus on the third term
 \bea
  && -\fr12\sqrt{h}\;h^{i'j'}\left(\pa_{i'} X^{u'} \pa_{j'} X^{v'} \eta_{u'v'}
      -\pa_{i'} X^{m'} \pa_{j'} X^{n'} \eta_{m'n'} \right)  \nn\\
 &&\left[\fr{}{}\right.
     \fr12\sqrt{h}\;h^{ij}\left( \frac{r^2}{{r_0}^2}
  - \frac{6\,{\left( X_0\cdot X \right) }^2}{{r_0}^4}
    \right)
     \left(  \pa_i X^u \pa_j X^v \eta_{uv}
            -\pa_i X^m \pa_j X^n \eta_{mn}\right)
    \nn\\
       &&-\fr1{(p^+)^2}\sqrt{h}h^{ij}\pa_i X^+\pa_j X^+
       \fr{1}{r_0^2}  \;\left(\fr{}{}\right.\fr{1}{192}(S\g^{au} S)( S\g^{au} S)
  -\fr{1}{192}(S\g^{ab} S)( S\g^{ab} S)
   \nn\\
 &&\quad  -\fr1{32}(S\g^{au} S)( S\g^{bu} S)
     \fr{X_0^aX_0^b}{r_0^2}
  +\fr1{32}(S\g^{ac} S)( S\g^{bc} S)
  \fr{X_0^aX_0^b}{r_0^2}
   \left.\fr{}{}\right)
   \left.\fr{}{}\right] \nn\\
 =&& \fr12\sqrt{h}\;h^{i'j'}\pa_{i'} X^{u'} \pa_{j'} X^{v'} \eta_{u'v'}
    \nn\\
       &&\fr1{(p^+)^2}\sqrt{h}h^{ij}\pa_i X^+\pa_j X^+
       \fr{1}{r_0^2}  \;\left(\fr{}{}\right.\fr{1}{192}(S\g^{au} S)( S\g^{au} S)
  -\fr{1}{192}(S\g^{ab} S)( S\g^{ab} S)
   \nn\\
 &&\quad  -\fr1{32}(S\g^{au} S)( S\g^{bu} S)
     \fr{X_0^aX_0^b}{r_0^2}
  +\fr1{32}(S\g^{ac} S)( S\g^{bc} S)
  \fr{X_0^aX_0^b}{r_0^2}
   \left.\fr{}{}\right) \nn\\
 \eea
All the other terms vanish due to dimensional regularization.
Together with the four external vector vertex operators the
correlator has the form of
 \bea
 && <V^{u_1}_v(x_1) V^{u_2}_v(x_2) V^{u_3}_v(x_3) V^{u_4}_v(x_4)
  \pa_{i'} X^{u'} \pa_{j'} X^{v'} \eta_{u'v'}
 (S\g^{AB} S)(S\g^{CD}S)>\nn\\
=&& <
+l^8R^{u_1v_1}k_1^{v_1}R^{u_2v_2}k_2^{v_2}R^{u_3v_3}k_3^{v_3}R^{u_4v_4}k_4^{v_4}\nn\\
&&+l^4\left[\dot{X}^{u_1}\dot{X}^{u_2}R^{u_3v_3}k_3^{v_3}R^{u_4v_4}k_4^{v_4}
  +\dot{X}^{u_3}\dot{X}^{u_4}R^{u_1v_1}k_1^{v_1}R^{u_2v_2}k_2^{v_2}\right.\nn\\
 && \left.\;\;\;\;+\dot{X}^{u_1}\dot{X}^{u_4}R^{u_2v_2}k_2^{v_2}R^{u_3v_3}k_3^{v_3}
  +\dot{X}^{u_1}\dot{X}^{u_3}R^{u_2v_2}k_2^{v_2}R^{u_4v_4}k_4^{v_4}\right.\nn\\
  &&\left. \;\;\;\;+\dot{X}^{u_2}\dot{X}^{u_3}R^{u_1v_1}k_1^{v_1}R^{u_4v_4}k_4^{v_4}
  +\dot{X}^{u_2}\dot{X}^{u_4}R^{u_1v_1}k_1^{v_1}R^{u_3v_3}k_3^{v_3}\right]
  \nn\\
 &&-l^6\left[\dot{X}^{u_1}R^{u_2v_2}k_2^{v_2}R^{u_3v_3}k_3^{v_3}R^{u_4v_4}k_4^{v_4}
        +\dot{X}^{u_2}R^{u_1v_1}k_1^{v_1}R^{u_3v_3}k_3^{v_3}R^{u_4v_4}k_4^{v_4}
        \right. \nn\\
&&\left.\;\;\;\;+\dot{X}^{u_3}R^{u_1v_1}k_1^{v_1}R^{u_2v_2}k_2^{v_2}R^{u_4v_4}k_4^{v_4}
        +\dot{X}^{u_4}R^{u_1v_1}k_1^{v_1}R^{u_2v_2}k_2^{v_2}R^{u_3v_3}k_3^{v_3}
        \right]\nn\\
   &&\pa_{i'} X^{u'} \pa_{j'} X^{v'} \eta_{u'v'}
 (S\g^{AB} S)(S\g^{CD}S)>
 \eea
Consider the 8S terms,
 \bea
&&+l^4\left[\dot{X}^{u_1}\dot{X}^{u_2}R^{u_3v_3}k_3^{v_3}R^{u_4v_4}k_4^{v_4}
  +\dot{X}^{u_3}\dot{X}^{u_4}R^{u_1v_1}k_1^{v_1}R^{u_2v_2}k_2^{v_2}\right.\nn\\
 && \left.\;\;\;\;+\dot{X}^{u_1}\dot{X}^{u_4}R^{u_2v_2}k_2^{v_2}R^{u_3v_3}k_3^{v_3}
  +\dot{X}^{u_1}\dot{X}^{u_3}R^{u_2v_2}k_2^{v_2}R^{u_4v_4}k_4^{v_4}\right.\nn\\
  &&\left. \;\;\;\;+\dot{X}^{u_2}\dot{X}^{u_3}R^{u_1v_1}k_1^{v_1}R^{u_4v_4}k_4^{v_4}
  +\dot{X}^{u_2}\dot{X}^{u_4}R^{u_1v_1}k_1^{v_1}R^{u_3v_3}k_3^{v_3}\right]
      \nn\\
   &&\pa_{i'} X^{u'} \pa_{j'} X^{v'} \eta_{u'v'}
 (S\g^{AB} S)(S\g^{CD}S)>
 \eea
Since $<RRSSSS>$-terms start with $\fr{1}{x_1^2}$ order, the
X-correlators must have $O(1)$-order terms. The leading order comes
when each $\pa_{i'} X^{u'}$ contracts with $e^{ik\cdot X}$. This
means that the above correlators are proportional to the
corresponding part of $<VVVVSSSS>$, which vanishes. As a matter of
fact this argument applies to the other cases i.e., $<RRRSSSS>$ and
$<RRRRSSSS>$. With these one gets
 \bea
 && <V^{u_1}_v V^{u_2}_v V^{u_3}_v V^{u_4}_v
  \pa_{i'} X^{u'} \pa_{j'} X^{u'}
 \;S\g^{AB} S\;\;S\g^{CD}S>
 = 0
 \eea
Therefore the third term entirely vanishes: the equation,
(\ref{VqVq-vec}), simplifies to
 \bea
 (V_{G,q})^2  =&&\fr{q^2}{4}\int  \int  \;\sqrt{h}\;h^{i'j'}
                 \;\sqrt{h}\;h^{ij}\left[\fr{}{}\right.
  +\fr14  \left(\pa_{i'} X^{u'} \pa_{j'} X^{u'}\right)
         \left(\pa_i X^u \pa_j X^u\right)\nn\\
   &&+4\frac{\,X_0^{l'} X^{l'}}{{r_0}^2}\frac{\,X_0^l X^l}{{r_0}^2}
     (\pa_{i'} X^{u'} \pa_{j'} X^{u'}
          \pa_i X^u \pa_j X^u
      + \pa_{i'} X^{m'} \pa_{j'} X^{m'}
      \pa_i X^m \pa_j X^m )\left.\fr{}{}\right]\nn\\
      \label{VqVqfinal2}
  \eea
  where the terms that vanish due to dimensional regularization
   have been dropped. The second term in (\ref{VqVqfinal2}) drops
   as one can see as follows. The second term in parentheses can
   be re-written,
 \bea
 && <X^{l'}\pa_{j'} X^{n'}\pa_{j'} X^{n'}\;X^l  \pa_i X^m \pa_i
 X^m>
 =-<\pa_{j'}X^{l'} X^{n'}\pa_{j'} X^{n'}\;X^l  \pa_i X^m \pa_i X^m>\nn\\
 =&&\fr12<\pa'^2X^{l'} X^{n'} X^{n'}\;X^l  \pa_i X^m \pa_i X^m>
 \eea
 which vanishes due to dimensional regularization. Consider the second term
 \bea
 &&   <X^{l'} X^l\;\pa_i X^u \pa_i X^u \;
           \pa_{j'} X^{u'} \pa_{j'} X^{u'}>\nn\\
= && -<\pa_i X^l X^{l'} X^u\pa_i X^u \;\pa_{j'} X^{u'}
   \pa_{j'}X^{u'}>
    - <X^{l'} X^l\; X^u \pa^2 X^u \;\pa_{j'} X^{u'} \pa_{j'} X^{u'}>\nn\\
= && -<\pa_i X^l X^{l'} X^u\pa_i X^u \;\pa_{j'} X^{u'}
\pa_{j'}X^{u'}>
    +<\pa_{j'} X^{l'} X^l\; X^u \pa^2 X^u \; X^{u'} \pa_{j'}
    X^{u'}>\nn\\
   &&  +<X^{l'} X^l\; X^u \pa^2 X^u \; X^{u'} \pa'^2 X^{u'}>
  \eea
The third term generates $\d(y-x_1)\d(y'-x_1)$ so it vanishes due to
dimensional regularization:
 \bea
= && -\pa_i X^l X^{l'} X^u\pa_i X^u \;\pa_{j'} X^{u'} \pa_{j'}X^{u'}
    +\pa_{j'} X^{l'} X^l\; X^u \pa^2 X^u \; X^{u'} \pa_{j'}
    X^{u'}\nn\\
 = &&\fr12\pa^2 X^l X^{l'} X^u X^u \;\pa_{j'} X^{u'} \pa_{j'}X^{u'}
  -\fr12\pa'^2 X^{l'} X^l\; X^u \pa^2 X^u \; X^{u'}X^{u'}\nn\\
 =&&\fr12 X^u X^u \;\pa_{j} X^{v} \pa_{j}X^{v}
    -\fr12  X^u \pa^2 X^u \; X^{v}X^{v}\nn\\
 =&& X^u X^u \pa_{j} X^{v} \pa_{j}X^{v}
    + X^u \pa_i X^u \;\pa_i X^{v}X^{v}
  \eea
Consider this together with VVVV. The RRRR-type terms do not
contribute due to momentum conservation. The computations with XXXX,
XRRR, XXRR have been carried out using Mathematica. The summary of
the results is as follows. The XXRR terms vanish either because they
are subleading in $\fr{1}{x_1}$ or they contain overall $k_1^2$.
Next, the correlators with the XRRR-type terms,
 \bea
 &&<-l^6\left[\dot{X}^{u_1}R^{u_2v_2}k_2^{v_2}R^{u_3v_3}k_3^{v_3}R^{u_4v_4}k_4^{v_4}
        +\dot{X}^{u_2}R^{u_1v_1}k_1^{v_1}R^{u_3v_3}k_3^{v_3}R^{u_4v_4}k_4^{v_4}
        \right. \nn\\
&&\left.\;\;\;\;+\dot{X}^{u_3}R^{u_1v_1}k_1^{v_1}R^{u_2v_2}k_2^{v_2}R^{u_4v_4}k_4^{v_4}
        +\dot{X}^{u_4}R^{u_1v_1}k_1^{v_1}R^{u_2v_2}k_2^{v_2}R^{u_3v_3}k_3^{v_3}
        \right]\nn\\
       && (X^u X^u \;\pa_{j} X^{v} \pa_{j}X^{v} + X^u \pa_i X^u \;\pa_i
       X^{v}X^{v})>
 \eea
When RRR contains $R^{u_1v_1}(x_1)$ it is already $\fr{1}{x_1^2}$
-order. The remaining $X$ correlators produce terms with
$O\left(\fr{1}{x_1}\right)$. As for the terms with RRR that do not
include $R^{u_1v_1}(x_1)$, the leading order contributions come from
the following $X$-contraction
 \bea
 &&<\dot{X}^{u_1}\;(X^u X^u \;\pa_{j} X^{v} \pa_{j}X^{v}
    + X^u \pa_i X^u \;\pa_i X^{v}X^{v})>
 \eea
All the terms of order $\fr{1}{x_1^2}$ contain at least one overall
factor of $k_1^2$, therefore they vanish. Finally consider the
correlators with XXXX. The $<XXXX\;(X^u X^u \;\pa_{j} X^{v}
\pa_{j}X^{v}>$-correlator comes with $k_1^2=0$. The $<XXXX\;(X^u
\pa_i X^u \;\pa_{i} X^{v} X^{v}>$-correlator turns out to contain
overall $\z_1\cdot k_1$, therefore vanishes: one gets
 \bea
 (V_{G,q})^2 &\Rightarrow &\fr{q^2}{16}\int \int   \;\sqrt{h}\;h^{i'j'}
                 \;\sqrt{h}\;h^{ij}
   \left(\pa_{i'} X^{u'} \pa_{j'} X^{u'} \right)
     \left(\pa_i X^u \pa_j X^u\right)\nn\\
  &\Rightarrow &\fr{q^2}{8}\int \;\sqrt{h}\;h^{ij}
     \left(\pa_i X^u \pa_j X^u\right)
     \label{VqVqfinalone}
 \eea

\subsection{Four scalar scattering}

Similar steps are involved for the four scalar correlators. The
explicit form of product of the four scalar vertex operators is
 \bea
&& V^{m_1}_s(x_1) V^{m_2}_s(x_2) V^{m_3}_s(x_3) V^{m_4}_s(x_4) \nn\\
&& =X'^{m_1}X'^{m_2}X'^{m_3}X'^{m_4}
+l^8R^{m_1v_1}k_1^{v_1}R^{m_2v_2}k_2^{v_2}R^{m_3v_3}k_3^{v_3}R^{m_4v_4}k_4^{v_4}\nn\\
&&-l^2\left[X'^{m_1}X'^{m_2}X'^{m_3}R^{m_4v_4}k_4^{v_4}
  +X'^{m_1}X'^{m_2}X'^{m_4}R^{m_3v_3}k_3^{v_3}\right.\nn\\
&&\left.\;\;\;\;+X'^{m_1}X'^{m_3}X'^{m_4}R^{m_2v_2}k_2^{v_2}
   +X'^{m_2}X'^{m_3}X'^{m_4}R^{m_1v_1}k_1^{v_1}\right]\nn\\
&&+l^4\left[X'^{m_1}X'^{m_2}R^{m_3v_3}k_3^{v_3}R^{m_4v_4}k_4^{v_4}
  +X'^{m_3}X'^{m_4}R^{m_1v_1}k_1^{v_1}R^{m_2v_2}k_2^{v_2}\right.\nn\\
 && \left.\;\;\;\;+X'^{m_1}X'^{m_4}R^{m_2v_2}k_2^{v_2}R^{m_3v_3}k_3^{v_3}
  +X'^{m_1}X'^{m_3}R^{m_2v_2}k_2^{v_2}R^{m_4v_4}k_4^{v_4}\right.\nn\\
  &&\left. \;\;\;\;+X'^{m_2}X'^{m_3}R^{m_1v_1}k_1^{v_1}R^{m_4v_4}k_4^{v_4}
  +X'^{m_2}X'^{m_4}R^{m_1v_1}k_1^{v_1}R^{m_3v_3}k_3^{v_3}\right]
  \nn\\
 &&-l^6\left[X'^{m_1}R^{m_2v_2}k_2^{v_2}R^{m_3v_3}k_3^{v_3}R^{m_4v_4}k_4^{v_4}
        +X'^{m_2}R^{m_1v_1}k_1^{v_1}R^{m_3v_3}k_3^{v_3}R^{m_4v_4}k_4^{v_4}
        \right. \nn\\
&&\left.\;\;\;\;+X'^{m_3}R^{m_1v_1}k_1^{v_1}R^{m_2v_2}k_2^{v_2}R^{m_4v_4}k_4^{v_4}
        +X'^{m_4}R^{m_1v_1}k_1^{v_1}R^{m_2v_2}k_2^{v_2}R^{m_3v_3}k_3^{v_3}
        \right] \label{4s}
 \eea

\subsubsection{$V_{G,q^2}$ contribution}

The operator takes the form of

 \bea
 V_{G,q^2}
 =&&{q^2}\int  \;(-)\fr{1}2\sqrt{h}\;h^{ij}
     \left(\pa_i X^u \pa_j X^v \eta_{uv}
     \left[  \frac{3}{8}
   - \frac{3\,X_0\cdot X}{{r_0}^2}
       +  \frac{-3\,r^2}{2\,{r_0}^2}
       + \frac{15\,{\left( X_0\cdot X \right) }^2}{{r_0}^4}
         \right] \right.\nn\\
   &&\left. \hspace{1.5in}
     +\pa_i X^m \pa_j X^n \eta_{mn}
     \left[   -\frac{1}{8}+\frac{X_0.X }{{r_0}^2}
  +\frac{r^2}{2 {r_0}^2}  -\frac{5(X_0.X)^2}{{r_0}^4}
  \right]
     \right)
  \nn\\
 && +\fr{i}{8p^+r_0^2} (\sqrt{h}h^{ij}-\ve^{ij})\pa_i X^+
     \left[\fr{}{}\right.
    7 \pa_jX^u(X+X_0)^m\; (S\g^{um}S)
        -5\pa_jX^m (X+X_0)^n\;(S\g^{mn}S)
       \left.\fr{}{}\right]\nn\\
       &&+\fr1{4(p^+)^2}\sqrt{h}h^{ij}\pa_i X^+\pa_j X^+
       \frac{1}{r_0^2}\;\left[\fr{}{}\right.
  -\fr{17}{96}(S\g^{uv} S)( S\g^{uv} S)
    +\fr{45}{48}(S\g^{au} S)( S\g^{au} S)\nn\\
  && \quad\quad
   -\fr1{6}(S\g^{ab} S)( S\g^{ab} S)
  +\fr{19}{48}(S\g^{au} S)( S\g^{bu} S)
       \frac{X_0^aX_0^b}{r_0^2}
    +\fr{35}{24}(S\g^{ac} S)( S\g^{bc} S)
     \frac{X_0^aX_0^b}{r_0^2}
  \left.\fr{}{}\right]
  \label{vgqq-scalar}\nn\\
  \eea
The reason for dropping $\fr{1}{p^+}$-terms is that these
correlators already appeared in the one-loop analysis
\cite{Park:2008fp}. Let's briefly review the discussion. The
vanishing of $<VVVV\pa X^m X_0^n S\g^{um}S>$ was explicitly
discussed in \cite{Park:2008fp}. Consider $\pa X^u X_0^m S\g^{um}S$
 \bea
&& <V^{m_1}_s(x_1) V^{m_2}_s(x_2) V^{m_3}_s(x_3) V^{m_4}_s(x_4)\;
   \pa X^u X_0^m S\g^{um}S> \nn\\
&& =<l^8R^{m_1v_1}k_1^{v_1}R^{m_2v_2}k_2^{v_2}R^{m_3v_3}k_3^{v_3}R^{m_4v_4}k_4^{v_4}\nn\\
&&+l^4\left[X'^{m_1}X'^{m_2}R^{m_3v_3}k_3^{v_3}R^{m_4v_4}k_4^{v_4}
  +X'^{m_3}X'^{m_4}R^{m_1v_1}k_1^{v_1}R^{m_2v_2}k_2^{v_2}\right.\nn\\
 && \left.\;\;\;\;+X'^{m_1}X'^{m_4}R^{m_2v_2}k_2^{v_2}R^{m_3v_3}k_3^{v_3}
  +X'^{m_1}X'^{m_3}R^{m_2v_2}k_2^{v_2}R^{m_4v_4}k_4^{v_4}\right.\nn\\
  &&\left. \;\;\;\;+X'^{m_2}X'^{m_3}R^{m_1v_1}k_1^{v_1}R^{m_4v_4}k_4^{v_4}
  +X'^{m_2}X'^{m_4}R^{m_1v_1}k_1^{v_1}R^{m_3v_3}k_3^{v_3}\right]
  \nn\\
   && \quad\;\;    \pa X^u  S\g^{um}S>
 \eea
The $RR$ terms vanish due to the index structure. One can easily see
this just by explicitly considering  the index structure of
$<RRSS>$. Consider the $RRRR$ term. All the terms in the resulting
computation contain traces of either three or five
$\g^{mu}$-matrices. When the trace is expressed in terms of the
product of deltas each term is bound to contain at least one
$\d_{mu}$, therefore vanishes. Similarly XXRR terms vanish. Consider
$\pa X^m X^n S\g^{mn}S$. Among (\ref{4s}) terms only XXRR can give a
potentially non-zero result
 \bea
 &&l^4<\left[X'^{m_1}X'^{m_2}R^{m_3v_3}k_3^{v_3}R^{m_4v_4}k_4^{v_4}
  +X'^{m_3}X'^{m_4}R^{m_1v_1}k_1^{v_1}R^{m_2v_2}k_2^{v_2}\right.\nn\\
 && \left.\;\;\;\;+X'^{m_1}X'^{m_4}R^{m_2v_2}k_2^{v_2}R^{m_3v_3}k_3^{v_3}
  +X'^{m_1}X'^{m_3}R^{m_2v_2}k_2^{v_2}R^{m_4v_4}k_4^{v_4}\right.\nn\\
  &&\left. \;\;\;\;+X'^{m_2}X'^{m_3}R^{m_1v_1}k_1^{v_1}R^{m_4v_4}k_4^{v_4}
  +X'^{m_2}X'^{m_4}R^{m_1v_1}k_1^{v_1}R^{m_3v_3}k_3^{v_3}\right]\nn\\
 &&\quad\;\;\pa X^m X^n S\g^{mn}S>\label{pzxmxnsgmns}
  \eea
Explicit computation shows that these terms are subleading in $x_1$,
i.e., $O\left(\fr{1}{x_1^3}\right)$. One can similarly show that the
$\pa X^u X^m S\g^{um}S$-correlator vanishes. As in the vector
scattering case one can show that $<V_sV_sV_sV_sSSSS>=0$ and from
the one-loop analysis \cite{Park:2008fp} we know that
$<V_sV_sV_sV_s\;\pa_i X^u \pa_j X^v \eta_{uv}>=0$. With these
results, (\ref{vgqq-scalar}) can be re-written as
 \bea
 V_{G,q^2}
 =&&{q^2}\int  \;(-)\fr{1}2\sqrt{h}\;h^{ij}
  \nn\\
 &&\left(\pa_i X^u \pa_j X^u
     \left[
   - \frac{3\,X_0\cdot X}{{r_0}^2}
       +  \frac{-3\,r^2}{2\,{r_0}^2}
       + \frac{15\,{\left( X_0\cdot X \right) }^2}{{r_0}^4}
     \right] \right.\nn\\
   &&\left.
     +\pa_i X^m \pa_j X^m
     \left[   -\frac{1}{8}+\frac{X_0.X }{{r_0}^2}
  +\frac{r^2}{2 {r_0}^2}
   -\frac{5(X_0.X)^2}{{r_0}^4}\right) \right]
 \eea
The first term goes as follows. Each $\pa_i X^u$ must contract with
$e^{ik\cdot X}$. The leading term comes with $k_1^2$, therefore
vanishes. The second term vanishes as follows. The $\pa_i X^m \pa_j
X^m (X_0.X )$-term appeared in the one-loop as well. It vanishes due
to the index structure. Finally the only non-trivial piece of the
quartic term has the form of
 \bea
   <X'^{m_1}X'^{m_2}X'^{m_3}X'^{m_4}\;
   (\pa_i X^m \pa_j X^m X^pX^q)>
 \eea
One can check that it only produces terms of $\fr{1}{x_1^3}$ or
higher, hence no contribution. Therefore we prove the third equation
of (\ref{correlator-summary}),
 \bea
 V_{G,q^2}
 \Rightarrow \fr{q^2}{16}\int  \;\sqrt{h}\;h^{ij}
      \pa_i X^m \pa_j X^m
 \eea

\subsubsection{ $V_{G,q}^2$ contribution }
Tailoring (\ref{vqmother}) to the given order, one gets
 \bea
 V_{G,q}^2 =&&{q^2}\int \int  \;\fr{1}4\sqrt{h}\;h^{ij}
                 \sqrt{h}\;h^{i'j'}\nn\\
 &&\left[\fr{}{}\right. -\frac{1}{2} \pa_i X^u \pa_j X^v \eta_{uv}
           +\frac{1}{2}\pa_i X^m \pa_j X^n \eta_{mn}\nn\\
           &&+\pa_i X^u \pa_j X^v \eta_{uv}
           \frac{2\,X_0\cdot X}{{r_0}^2}
            -\pa_i X^m \pa_j X^n \eta_{mn}
          \frac{2 X_0.X }{{r_0}^2}\nn\\
    && +\pa_i X^u \pa_j X^v \eta_{uv}
     \left( \frac{r^2}{{r_0}^2}
  - \frac{6\,{\left( X_0\cdot X \right) }^2}{{r_0}^4}
    \right)
      \nn\\
   &&+\pa_i X^m \pa_j X^n \eta_{mn}
       \left(\frac{6(X_0.X)^2}{{r_0}^4}-\frac{r^2}{{r_0}^2}\right)
     \left.\fr{}{}\right]^2
  \nn\\
 && +\fr1{4(p^+)^2} \left[\fr{}{}\right.
   -{i} (\sqrt{h}h^{ij}-\ve^{ij})\pa_i X^+
      \pa_jX^uX_0^m\; (S\g^{um}S)
     \frac{1}{r_0^2}
        \nn\\
  && \quad\quad\quad   +{i} (\sqrt{h}h^{ij}-\ve^{ij})\pa_i X^+
       \pa_jX^m X_0^n\;(S\g^{mn}S)
   \frac{1}{r_0^2}
       \left.\fr{}{}\right]^2\nn\\
 && +\fr1{p^+} \left[\fr{}{}\right.
   \fr{i}4 (\sqrt{h}h^{i'j'}-\ve^{i'j'})\pa_{i'} X^+
      \pa_{j'}X^{u'}(X+X_0)^{m'}\; (S\g^{u'm'}S)
     \left(  -\frac{4}{r_0^2}+\frac{24 X_0.X }{r_0^4} \right)
        \nn\\
  && \quad\quad\quad   -\fr{i}4 (\sqrt{h}h^{i'j'}-\ve^{i'j'})\pa_{i'} X^+
       \pa_{j'}X^{m'} (X+X_0)^{n'}\;(S\g^{m'n'}S)
  \left( -\frac{4}{r_0^2}+\frac{24X_0.X }{r_0^4}\right)
       \left.\fr{}{}\right]\nn\\
  &&\sqrt{h}\;h^{ij}\left[\pa_i X^u \pa_j X^v \eta_{uv}
      \left( - \frac{1}{2}
  + \frac{2\,X_0\cdot X}{{r_0}^2}  \right)
     +\pa_i X^m \pa_j X^n \eta_{mn}
       \left(\frac{1}{2}-\frac{2 X_0.X }{{r_0}^2}\right)
     \right]\nn\\ \label{Vsqqsimple}
 \eea
where we have used
 \bea
 && <V^{m_1}_s V^{m_2}_s V^{m_3}_s V^{m_4}_s
  \pa_{i'} X^{m'} \pa_{j'} X^{m'}
 (S\g^{AB} S)(S\g^{CD}S)>
 = 0
 \eea
and
 \bea
 && <V^{m_1}_s V^{m_2}_s V^{m_3}_s V^{m_4}_s
  \pa_{i'} X^{u'} \pa_{j'} X^{v'} \eta_{u'v'}
 (S\g^{AB} S)(S\g^{CD}S)>=0
 \eea
 Let's work out one of the last terms with the factor,
 \bea
 && \pa_{j'}X^{u'}(X+X_0)^{m'}\; (S\g^{u'm'}S)
     \left(  -\frac{4}{r_0^2}+\frac{24 X_0^{p'}X^{p'} }{r_0^4} \right)\nn\\
   &&\left[\pa_i X^u \pa_j X^v \eta_{uv}
      \left( - \frac{1}{2}
  + \frac{2\,X_0\cdot X}{{r_0}^2}  \right)
     +\pa_i X^m \pa_j X^n \eta_{mn}
       \left(\frac{1}{2}-\frac{2 X_0.X t}{{r_0}^2}\right)
     \right]\nn\\
 =&& \pa_{j'}X^{u'}X^{m'}\; (S\g^{u'm'}S)
     \left(  -\frac{4}{r_0^2} \right)
    \left[\pa_i X^u \pa_j X^v \eta_{uv}
      \left( - \frac{1}{2}
  \right)
     +\pa_i X^m \pa_j X^n \eta_{mn}
       \left(\frac{1}{2}\right)
     \right]\nn\\
  &&+ \pa_{j'}X^{u'}X_0^{m'}\; (S\g^{u'm'}S)
     \left(  -\frac{4}{r_0^2}+\frac{24 X_0^{p'}X^{p'} }{r_0^4} \right)\nn\\
     &&\left[\pa_i X^u \pa_j X^v \eta_{uv}
      \left( - \frac{1}{2}
  + \frac{2\,t\,X_0\cdot X}{{r_0}^2}  \right)
     +\pa_i X^m \pa_j X^n \eta_{mn}
       \left(\frac{1}{2}-\frac{2 X_0.X t}{{r_0}^2}\right)
     \right] \label{ssterm}
 \eea
Consider the first line. With rather lengthy algebra, one can show,
after the Wick rotation, that
 \bea
&& <V^{m_1}_s(x_1) V^{m_2}_s(x_2) V^{m_3}_s(x_3) V^{m_4}_s(x_4)
    \pa_{i'} X^{m'} \pa_{i'}X^{m'}
       \pa_jX^u X^m\;(S\g^{um}S)>\nn\\
  = &&2stu
  \left[\fr{}{}\right.
     \fr{\xi_1\cdot\xi_4\xi_2\cdot\xi_3}{1+\a't}
     + \fr{{\xi_1\cdot\xi_3\xi_2\cdot\xi_4}}{1+\a'u}
 +\fr{\xi_1\cdot\xi_2\xi_3\cdot\xi_4}{1+\a's}
    \left.\fr{}{}\right]
   \eea
What happens is that this contribution is precisely canceled by the
other term in the first line. In other words,
 \bea
  <&&V^{m_1}_s(x_1) V^{m_2}_s(x_2) V^{m_3}_s(x_3)V^{m_4}_s(x_4)\;\nn\\
   && \pa_{j'}X^{u'}X^{m'}\; (S\g^{u'm'}S)
    \left[-\pa_i X^u \pa_j X^v \eta_{uv}
     +\pa_i X^m \pa_j X^n \eta_{mn}
     \right]>=0 \label{3looprelated}
  \eea
This result has some implication for the three-loop computation as
we will discuss in the conclusion. The second term of (\ref{ssterm})
is
 \bea
  && \pa_{j'}X^{u'}X_0^{m'}\; (S\g^{u'm'}S)
     \left(  -\frac{4}{r_0^2}+\frac{24 X_0^{p'}X^{p'} t}{r_0^4} \right)\nn\\
     &&\left[\pa_i X^u \pa_j X^v \eta_{uv}
      \left( -\left( \frac{1}{2} \right)
  + \frac{2\,t\,X_0\cdot X}{{r_0}^2}  \right)
     +\pa_i X^m \pa_j X^n \eta_{mn}
       \left(\frac{1}{2}-\frac{2 X_0.X t}{{r_0}^2}\right)
     \right]\nn\\
  =&&   -\frac{2}{r_0^2}\pa_{j'}X^{u'}X_0^{m'}\; (S\g^{u'm'}S)
      \left[-\pa_i X^u \pa_j X^v \eta_{uv}
             +\pa_i X^m \pa_j X^n \eta_{mn}
     \right]\nn\\
  &&-\frac{8}{r_0^2}\pa_{j'}X^{u'}X_0^{m'}\; (S\g^{u'm'}S)
     \frac{\,X_0\cdot X}{{r_0}^2}
     \left[\pa_i X^u \pa_j X^v \eta_{uv}
     -\pa_i X^m \pa_j X^n \eta_{mn}
     \right]\nn\\
  && +12\pa_{j'}X^{u'}X_0^{m'}\; (S\g^{u'm'}S)
     \left(  \frac{ X_0^{p'}X^{p'} }{r_0^4} \right)
     \left[-\pa_i X^u \pa_j X^v \eta_{uv}
     +\pa_i X^m \pa_j X^n \eta_{mn}
     \right]\nn\\
  = &&-\frac{8}{r_0^2}\pa_{j'}X^{u'}X_0^{m'}\; (S\g^{u'm'}S)
     \frac{\,X_0\cdot X}{{r_0}^2}
     \left[\pa_i X^u \pa_j X^v \eta_{uv}
     -\pa_i X^m \pa_j X^n \eta_{mn}
     \right]\nn\\
  && +12\pa_{j'}X^{u'}X_0^{m'}\; (S\g^{u'm'}S)
     \left(  \frac{ X_0^{p'}X^{p'} }{r_0^4} \right)
     \left[-\pa_i X^u \pa_j X^v \eta_{uv}
     +\pa_i X^m \pa_j X^n \eta_{mn}
     \right]
 \eea
where the last equality follows from the fact that the second term
vanishes due to the index structure. The vanishing of the terms in
the last line goes almost identical to one of the previous
calculations that leads to (\ref{3looprelated}). The above equation
becomes
 \bea
  = &&-\frac{8}{r_0^2}\pa_{j'}X^{u'}X_0^{m'}\; (S\g^{u'm'}S)
     \frac{\,X_0\cdot X}{{r_0}^2}
     \left[\pa_i X^u \pa_j X^v \eta_{uv}
     -\pa_i X^m \pa_j X^n \eta_{mn}
     \right]
 \eea
  Correlators with XXRR and RRRR  types of terms vanish since there are
  only odd number of $X^m$'s. First the correlator involving the XXXR terms,
 \bea
  XXXR\;\pa_{j'}X^{u'} X^p
  [\pa_i X^u \pa_j X^v \eta_{uv}
     -\pa_i X^m \pa_j X^n \eta_{mn}](S\g^{u'm'}S)
 \eea
 Partially integrate $\pa_{j'}X^{u'}$ so that the derivative now acts on
 $(S\g^{u'm'}S)$. It should contract only with $R^{m_1v_1}$. The $\pa' S'$
 contraction gives a delta function. The  other $S$ will vanish due to dimensional
  regularization. Next the XRRR-correlator,
 \bea
  XRRR\;\pa_{j'}X^{u'} X^p
  [\pa_i X^u \pa_j X^v \eta_{uv}
     -\pa_i X^m \pa_j X^n \eta_{mn}](S\g^{u'm'}S)
 \eea
 Again partially integrate $\pa_{j'}X^{u'}$ so that the derivative now acts on
 $(S\g^{u'm'}S)$. It is easy to check that both of these terms
  produce sub-leading terms in $\fr{1}{x_1}$, i.e.,
  $O\left(\fr{1}{x_1^3}\right)$. The third term in (\ref{ssterm}) can
  also be shown to vanish,
 \bea
<V_sV_sV_sV_s\left[\pa_i X^u \pa_j X^u
      \left( - \frac{1}{2}
  + \frac{2\,X_0\cdot X}{{r_0}^2}  \right)
     +\pa_i X^m \pa_j X^m
       \left(\frac{1}{2}-\frac{2 X_0.X }{{r_0}^2}\right)
     \right]>=0 \label{continue-app}
 \eea
 To prevent this section from becoming too lengthy we do not present
 the discussion for the rest of the terms in (\ref{Vsqqsimple}). One can
  show by following the similar steps that all the remaining terms
  in (\ref{Vsqqsimple}) vanish except one term:
 \bea
 V_{G,q}^2 &\Rightarrow &\fr{q^2}{16}\int \int \sqrt{h}\;h^{ij}
                  \sqrt{h}\;h^{i'j'}
           (\pa_{i'} X^{m'} \pa_{j'} X^{m'})
            (\pa_i X^m \pa_j X^m)\nn\\
      &\Rightarrow &\fr{q^2}{8}\int \;\sqrt{h}\;h^{ij}
     \left(\pa_i X^m \pa_j X^m\right)
  \label{VsqqsimplestFinal}
 \eea

\section{Conclusion}

In this paper, we have discussed the two-loop divergence
cancellation of the four vector and the four scalar scattering
amplitudes. At two-loop, the divergence vanishes.\footnote{ At least
the one that we have considered vanishes. The three-loop result will
most likely generate new forms of divergences. Any new form of
divergence cannot be cancelled by the counter-terms based on the
flat action. In this light, the three-loop result is expected to
completely rule out the flat action.} Compared with the one-loop
analysis, a new ingredient is the introduction of renormalization of
the external vertex operators. The divergences could be absorbed as
they have precisely the same forms as the tree-level ones. It is
necessary to consider some other scattering amplitudes in order to
complete the renormalization program in this paper. By doing so, it
should be possible to fix the various constants that appear in
(\ref{rnconst-rel1}), (\ref{rnconst-rel2}), (\ref{rnconst-rel3}),
(\ref{rnconst-rel4}) and (\ref{rnconst-rel5}). Some of these
equations may get modified by the consideration of $< VVVV
V_{G,g^2}>_{\mbox{1 loop}}$, which was noted in the footnote 7. We
will check in the near future whether the divergences from $< VVVV
V_{G,g^2}>_{\mbox{1 loop}}$ indeed preserve the consistency of the
proposal other than possibly modifying the constraint equations. It
will be interesting to compare those results with the wave function
renormalization of {N}=4 SYM. This will set the ground for the
computations of the open string corrections for the anomalous
dimensions of the SYM operators. The results may also be compared
with the results based on \cite{Park:2007ev}.

At the two-loop order the counter vertex operator contains a large
number of terms. As we have discussed in the main body, the majority
of these terms have turned out to vanish. All the non-vanishing
terms yield the same kinematic structure as that of the tree
diagrams. Because of this one may wonder whether it would be
possible at three-loop to have non-vanishing contributions that do
not have the same structures as the tree diagrams. As a matter of
fact, there are a few terms that have appeared at two-loop which
will re-appear at three-loop. For example, one of them is
 \bea
   && \pa_{j'}X^{u'}X^{m'}\; (S\g^{u'm'}S)
    \;\pa_i X^u \pa_j X^v \eta_{uv}
  \eea
At two-loop the contribution of this term gets cancelled by a
similar term
 \bea
   && \pa_{j'}X^{u'}X^{m'}\; (S\g^{u'm'}S)
    \;\pa_i X^m \pa_j X^n \eta_{mn}
  \eea
as seen in (\ref{3looprelated}). At three-loop, however, these two
terms emerge with coefficients of different magnitudes. Therefore we
expect that there will be a non-tree like divergence that is
provided by the counter vertex operator. This may be viewed as a
prediction that the three-loop diagrams will be divergent but not
finite. It will be a very interesting and crucial test of the
conjecture in \cite{Park:2008sg,Park:2008fp} to check whether the
picture holds at three-loop.\footnote{The two terms above are absent
in the flat space action. If they play a role in the three-loop
divergence cancellation as expected, it will be another argument in
addition to the one given in \cite{Park:2008fp} that the flat space
action is simply not suitable. } The three-loop computation seems to
be more within the range with the development of the pure spinor
formulation \cite{Berkovits:2005df}\cite{Berkovits:2000fe}. (See
also \cite{Lee:2006pa}\cite{Grassi:2001ug}\cite{Grassi:2004xr} for
related works.) Once one has a handle on the three loop
computations, one may be in a good position to concretely test  the
idea put forward in \cite{Park:1999xz} (and subsequent works)
concerning the potential roles of an open string in gauge/gravity
type conjectures in general. (Related ideas can be found, e.g., in
\cite{Kawai:2007ek}\cite{Azeyanagi:2008mi}\cite{DiVecchia:2005vm}.)

In the introduction, it was mentioned that with our set-up one may
not have to consider the non-planar graphs when one considers an
open string on a Dp-brane with $p<9$. The reason is that, for a
non-planar graph, at least one of the external states must leave the
brane at some point. This would be a violation of the boundary
conditions. This is a different situation from the space-time
filling brane where the end points will still move within the brane
since there is no leaving the brane. It will be interesting and
worthwhile to check explicitly what would go wrong mathematically in
the construction of such a process. (It might be that all such
amplitudes vanish.) If true (we believe that it is), one may say
that the planarity is built in the present frame-work but not
something that one imposes as an extra condition. It also has an
implication for T-duality. Naively one would expect to include
non-planar graphs since they are included in the D9-brane case.
Therefore this is an indication that care is needed when applying
T-duality. We will report on this issue and the ones given above in
the near future.

Finally we elaborate on the subtle issues with T-duality in addition
to the one just described. The issue will force us to face a
non-trivial question concerning how, with regard to the degrees of
freedom, one should formulate the string theory, open and/or closed.
Here we examine the problem from the angle of the current context
and point out a few things that may be useful to precisely pin-down
the problem. We also list future tasks that may lead to a satisfying
resolution on how to formulate the string theory.

It seems that we are dealing with two T-dualities depending on where
we apply it. The first is a T-duality that relates a D9 to a Dp with
$p<9$. The other is a "T-duality within the same Dp-brane theory".
(Below we explain what is meant by this.) To some extent, the
subtlety hinges on the set-up that we are using. Let's start with
the T-duality in the conventional set-up where both an open string
and a closed string are put on an equal footing. Although it may
sound straightforward, it is not the case once one considers loops
of the open strings that are moving on Dp-branes. We will name a
few. In this set-up one would consider the momentum modes along the
compactified directions and winding modes. Therefore there is no
loss of degrees of freedom. However, an explicit construction of the
open string scattering states requires them to have momentum
components solely along the longitudinal directions as shown in
\cite{Park:2007mc}. It is not clear to us how to reconcile or
whether it is even possible to do so. (Also in light of the open
string based framework which we discuss below, it is not clear
whether including those modes is justified. See the discussion
below.) We believe that a more serious issue exists within the
set-up itself. It is the redundancy issue. The issue is not limited
to the current context but much more general. In a certain
circumstance, the end points of an open string can stick together,
thereby converting it to a closed string. That seems to suggest
that, in principle, the resulting closed string should have a
representation in terms of the open string that one started with:
putting independent closed sting coordinates seems redundant.
Furthermore, independent closed string coordinates do not seem to
help when it comes to cancelling the open string divergences. A
closed string amplitude comes with the kinematic factor that is
distinct from that of an open string amplitude. As well known, one
has to "square" the latter to go to the former. We find obscure (and
could not find an explicit proof of) the statements in the
literatures that those divergence can be cancelled by closed string
amplitudes. If successful at the three-loop, one of the things that
the current program will bring as a by-product is a firmer
establishment of the open string based setup.

We turn to the open string based set-up that has served as a basis
for the rationale of our current and previous works. In this
frame-work, one tries to cancel the given loop divergence by
inserting a composite vertex operator. Suppose one starts with D9
and goes to a Dp-brane by T-duality. One then tries to relate the
loop amplitude of the D9 to that of the Dp. As stated above and in
the previous works, there is a loss of degrees of freedom since the
momentum can only be along the longitudinal directions. If one deals
with a corresponding situation in an ordinary quantum field theory,
it is obvious that one's attempt to relate the two results is not
justified, since the lower dimensional theory would be a
dimensionally {\em reduced} theory. In a string theory the fact that
is is not justified seems less obvious. First of all, the Dp-brane
theory is still a ten-dimensional theory. Secondly, the
simple-minded application of T-duality to the final form of the
D9-brane amplitude leads to the same Dp-brane amplitude that is
obtained by direct computation. We believe that it should be a
coincidence because the loss of degrees of freedom still remains
true. The reason behind the coincidence must be that the form of the
amplitude is highly constrained by various symmetries and the
dimensional analysis. The other T-duality i.e., the "T-duality
within the Dp-brane theory" seems safer. For example, one may be
able to deduce the scalar amplitude from the vector amplitude and
vice versa. Here those two multiplets are on an equal footing.
\footnote{The term, "T-duality within the same Dp", is used in this
sense: as in a usual quantum field theory, amplitudes of different
set of fields may sometimes be connected by a certain
transformation. If one traces its stringy origin, it is likely to
originate from the usual T-dualities combined in some way.} It might
not be completely subtlety-free, however, because one would have to
make sure, e.g., that T-duality commutes with dimensional
regularization. (In all of the examples that we have considered so
far T-duality gives the same results as those obtained by direct
computations.) In this work, therefore, we have explicitly and
directly computed all the correlators.

 \vspace{.5in}
\ni {\bf Acknowledgements}: I would like to acknowledge the KITP
scholar program
 through which I participated in the stimulating workshop on the
 pure spinor formulation and string field theory.

\newpage

\section*{Appendices}
\appendix

\renewcommand{\theequation}{A.\arabic{equation}}
\setcounter{equation}{0}

\section{Definitions of $V_v$ and $V_s$}

\ni For convenience we quote from \cite{Park:2007mc} the expressions
for the vector multiplet, $V_v$, (which was called $V_v$ in
\cite{Park:2007mc}) and the scalar multiplet, $V_s$.\footnote{Here
we only quote the bosonic vertex operators. The fermionic vertex
operators can be found in \cite{Park:2007mc}. The veertex operators
were obtained in the world-sheet strip. To compute correlators using
the Wick contraction technique as done in this work and the previous
works, one should go to the upper half-plane by the standard
conformal transformation. } Define $k^i=(k^u,0),\; \z^i=(\z^u,0)$
with $i=(u,m)$. The index $i$ here is not to be confused with the
world-sheet label used in the main body of the text. The bosonic
vector vertex operator is
 \bea
 V_{v}(\z,k)&=&
 (\z^uB_v^u-\z^-B_v^+)e^{ik\cdot X}
 \eea
where
 \bea
 B_v^+&=&p^+ \nn\\
 B_v^u&=& (\dot{X}^u-R_v^{uj}k^j)\nn\\
 F_v^{\dot{a}}&=&\fr{1}{\sqrt{2p^+}}[((\g^u)^T \dot{X}^uS)^{\dot{a}}
      -((\g^m)^T {{X^m}'}S)^{\dot{a}}
      +\fr13 :((\g^i)^TS)^{\dot{a}}R_v^{ij}:k^j]\nn\\
 F_v^a&=&  \sqrt{\fr{{p^+}}{2}}\;S^a
 \eea
where $R_v^{ij}=\fr14 S\g^{ij}S$. \ni For the scalar vertex
operator, we define $k^i=(k^u,0),\; \xi^i=(0,\xi^m)$:
 \bea
 V_{s}(\xi,k)&=&\xi\cdot B_s e^{ik\cdot X}=
 (\xi^mB_s^m)e^{ik\cdot X}
 \eea
where
 \bea
 B_s^m&=& ({X'}^m+R_s^{mj}k^j)\nn\\
 F_s^{\dot{a}}&=&\fr{1}{\sqrt{2p^+}}[((\g^u)^T \dot{X}^uS)^{\dot{a}}
 -((\g^m)^T {{X^m}'}S)^{\dot{a}}
        -\fr13 :((\g^i)^TS)^{\dot{a}}R_s^{ij}:k^j]\nn\\
 F_s^a&=&  -\sqrt{\fr{{p^+}}{2}}\;S^a
 \eea
 where $R_s^{ij}=\fr14 S\g^{ij}S=R_v^{ij}$.

\renewcommand{\theequation}{B.\arabic{equation}}
\setcounter{equation}{0}

\section{Identities}

For the computations in sec3, it is useful to note the several
identities. For example, one has
 \bea
  <R^{u_1v_1}R^{u_2v_2}>=
  -\fr{(\d_{u_1u_2}\d_{v_1v_2}-\d_{u_1v_2}\d_{u_2v_1})}{(x_1-x_2)^2}
 \eea
and
 \bea
&& <R^{u_1v_1}(x_1)R^{u_2v_2}(x_2)R^{u_3v_3}(x_3)> \nn\\
 =&&-\fr{1}{x_{12}x_{23}x_{13}}\;
 (\d_{u_2u_3}\d_{u_1v_2}\d_{v_1v_3}-\d_{u_2u_3}\d_{u_1v_3}\d_{v_1v_2}
  -\d_{u_2v_3}\d_{u_1v_2}\d_{v_1u_3}\nn\\
  &&\hspace{.8in} +\d_{u_2v_3}\d_{u_1u_3}\d_{v_1v_2}
  -\d_{u_3v_2}\d_{u_1u_2}\d_{v_1v_3}
  +\d_{u_3v_2}\d_{u_1v_3}\d_{u_2v_1}\nn\\
  &&\hspace{.8in}+\d_{v_2v_3}\d_{u_1u_2}\d_{v_1u_3}
  -\d_{v_2v_3}\d_{u_1u_3}\d_{u_2v_1})
 \eea
From these one can easily deduce
 \bea
 &&\Tr\; \g^{u_1v_1}\g^{u_2v_2}= -8(\d_{u_1u_2}\d_{v_1v_2}
  -\d_{u_1v_2}\d_{u_2v_1})
  \eea
and
 \bea
&& \Tr \left(\g^{u_1v_1}\g^{u_2v_2}\g^{u_3v_3}\right)\nn\\
 =&&8
 (\d_{u_2u_3}\d_{u_1v_2}\d_{v_1v_3}-\d_{u_2u_3}\d_{u_1v_3}\d_{v_1v_2}
  -\d_{u_2v_3}\d_{u_1v_2}\d_{v_1u_3}+\d_{u_2v_3}\d_{u_1u_3}\d_{v_1v_2}\nn\\
  &&
  -\d_{u_3v_2}\d_{u_1u_2}\d_{v_1v_3}
  +\d_{u_3v_2}\d_{u_1v_3}\d_{u_2v_1}+\d_{v_2v_3}\d_{u_1u_2}\d_{v_1u_3}
  -\d_{v_2v_3}\d_{u_1u_3}\d_{u_2v_1})
 \eea
The proof of the trace that involves four $\g$'s is more lengthy: it
yields\footnote
  {Note that the following identity is for 16 by 16 gamma matrices.
  For 8 by 8 matrices, a factor $\fr{1}{2^3}$ should appear
   instead of $\fr{1}{2^4}$ in the first line. }
  \bea
 &&\fr{1}{2^4}\;
  \Tr \g^{u_1v_1}\g^{u_2v_2}\g^{u_3v_3}\g^{u_4v_4}\nn\\
 =\quad &&\d_{u_1u_2}\d_{u_3u_4}
 (\d_{v_1v_2}\d_{v_3v_4}-\d_{v_1v_3}\d_{v_2v_4}+\d_{v_1v_4}\d_{v_2v_3})\nn\\
 +\!\!&&\d_{u_1u_3}\d_{u_2u_4}
 (-\d_{v_1v_2}\d_{v_3v_4}+\d_{v_1v_3}\d_{v_2v_4}-\d_{v_1v_4}\d_{v_2v_3})\nn\\
 +\!\!&& \d_{u_1u_4}\d_{u_2u_3}
 (\d_{v_1v_2}\d_{v_3v_4}-\d_{v_1v_3}\d_{v_2v_4}+\d_{v_1v_4}\d_{v_2v_3})
                     \nn\\
+&&\d_{u_1u_2}
 (\;\;-\d_{v_1v_2}\d_{u_3v_3}\d_{u_4v_4}-\d_{v_1v_2}\d_{u_3v_4}\d_{v_3u_4}
  +\d_{v_1u_3}\d_{v_2v_3}\d_{u_4v_4}-\d_{v_1u_3}\d_{v_2u_4}\d_{v_3v_4}\nn\\
  &&\hspace{.4in}+\d_{v_1u_3}\d_{v_2v_4}\d_{v_3u_4}-\d_{v_1v_3}\d_{v_2u_3}\d_{u_4v_4}
 +\d_{v_1v_3}\d_{v_2u_4}\d_{u_3v_4}+\d_{v_1u_4}\d_{v_2u_3}\d_{v_3v_4}\nn\\
 &&\hspace{.4in}-\d_{v_1u_4}\d_{v_2v_3}\d_{u_3v_4}+\d_{v_1u_4}\d_{v_2v_4}\d_{u_3v_3}
 -\d_{v_1v_4}\d_{v_2u_3}\d_{v_3u_4}-\d_{v_1v_4}\d_{v_2u_4}\d_{u_3v_3})
                 \nn\\
+&&\d_{u_1u_3}
 (\;\;-\d_{v_1u_2}\d_{v_2v_3}\d_{u_4v_4}+\d_{v_1u_2}\d_{v_2u_4}\d_{v_3v_4}
  -\d_{v_1u_2}\d_{v_2v_4}\d_{v_3u_4}+\d_{v_1v_2}\d_{u_2v_3}\d_{u_4v_4}\nn\\
  &&\hspace{.4in}+\d_{v_1v_2}\d_{u_2v_4}\d_{v_3u_4}-\d_{v_1v_3}\d_{u_2v_2}\d_{u_4v_4}
 -\d_{v_1v_3}\d_{u_2v_4}\d_{v_2u_4}+\d_{v_1u_4}\d_{u_2v_2}\d_{v_3v_4}\nn\\
 &&\hspace{.4in}-\d_{v_1u_4}\d_{u_2v_3}\d_{v_2v_4}+\d_{v_1u_4}\d_{u_2v_4}\d_{v_2v_3}
  -\d_{v_1v_4}\d_{u_2v_2}\d_{v_3u_4}+\d_{v_1v_4}\d_{u_2v_3}\d_{v_2u_4})
                 \nn\\
 +&&\d_{u_1u_4}
 (\;\;-\d_{v_1u_2}\d_{v_2u_3}\d_{v_3v_4}+\d_{v_1u_2}\d_{v_2v_3}\d_{u_3v_4}
  -\d_{v_1u_2}\d_{v_2v_4}\d_{u_3v_3}-\d_{v_1v_2}\d_{u_2v_3}\d_{u_3v_4}\nn\\
  &&\hspace{.4in}+\d_{v_1v_2}\d_{u_2v_4}\d_{u_3v_3}-\d_{v_1u_3}\d_{u_2v_2}\d_{v_3v_4}
 +\d_{v_1u_3}\d_{u_2v_3}\d_{v_2v_4}-\d_{v_1u_3}\d_{u_2v_4}\d_{v_2v_3}\nn\\
 &&\hspace{.4in}+\d_{v_1v_3}\d_{u_2v_2}\d_{u_3v_4}+\d_{v_1v_3}\d_{u_2v_4}\d_{v_2u_3}-
  \d_{v_1v_4}\d_{u_2v_2}\d_{u_3v_3}-\d_{v_1v_4}\d_{u_2v_3}\d_{v_2u_3})
                 )\nn\\
 +&&\d_{u_2u_3}
 (\;\;-\d_{u_1v_1}\d_{v_2v_3}\d_{u_4v_4}+\d_{u_1v_1}\d_{v_2u_4}\d_{v_3v_4}
  -\d_{u_1v_1}\d_{v_2v_4}\d_{v_3u_4}+\d_{u_1v_2}\d_{v_1v_3}\d_{u_4v_4}\nn\\
  &&\hspace{.4in}-\d_{u_1v_2}\d_{v_1u_4}\d_{v_3v_4}+\d_{u_1v_2}\d_{v_1v_4}\d_{v_3u_4}
 -\d_{u_1v_3}\d_{v_1v_2}\d_{u_4v_4}+\d_{u_1v_3}\d_{v_1u_4}\d_{v_2v_4}\nn\\
 &&\hspace{.4in}-\d_{u_1v_3}\d_{v_1v_4}\d_{v_2u_4}-\d_{u_1v_4}\d_{v_1v_2}\d_{v_3u_4}
  +\d_{u_1v_4}\d_{v_1v_3}\d_{v_2u_4}-\d_{u_1v_4}\d_{v_1u_4}\d_{v_2v_3})
                    \nn\\
 +&&\d_{u_2u_4}
 (\;\;-\d_{u_1v_1}\d_{v_2u_3}\d_{v_3v_4}+\d_{u_1v_1}\d_{v_2v_3}\d_{u_3v_4}
 - \d_{u_1v_1}\d_{v_2v_4}\d_{u_3v_3}+\d_{u_1v_2}\d_{v_1u_3}\d_{v_3v_4}\nn\\
  &&\hspace{.4in}-\d_{u_1v_2}\d_{v_1v_3}\d_{u_3v_4}+\d_{u_1v_2}\d_{v_1v_4}\d_{u_3v_3}
 +\d_{u_1v_3}\d_{v_1v_2}\d_{u_3v_4}-\d_{u_1v_3}\d_{v_1u_3}\d_{v_2v_4}\nn\\
 &&\hspace{.4in}+\d_{u_1v_3}\d_{v_1v_4}\d_{v_2u_3}-\d_{u_1v_4}\d_{v_1v_2}\d_{u_3v_3}
  +\d_{u_1v_4}\d_{v_1u_3}\d_{v_2v_3}-\d_{u_1v_4}\d_{v_1v_3}\d_{v_2u_3})
                        \nn\\
 +&&\d_{u_3u_4}
 (\;\;-\d_{u_1v_1}\d_{u_2v_2}\d_{v_3v_4}+\d_{u_1v_1}\d_{u_2v_3}\d_{v_2v_4}-
  \d_{u_1v_1}\d_{u_2v_4}\d_{v_2v_3}-\d_{u_1v_2}\d_{v_1u_2}\d_{v_3v_4}\nn\\
  &&\hspace{.4in}+\d_{u_1v_2}\d_{v_1v_3}\d_{u_2v_4}-\d_{u_1v_2}\d_{v_1v_4}\d_{u_2v_3}
 +\d_{u_1v_3}\d_{v_1u_2}\d_{v_2v_4}-\d_{u_1v_3}\d_{v_1v_2}\d_{u_2v_4}\nn\\
 &&\hspace{.4in}+\d_{u_1v_3}\d_{v_1v_4}\d_{u_2v_2}-\d_{u_1v_4}\d_{v_1u_2}\d_{v_2v_3}
  +\d_{u_1v_4}\d_{v_1v_2}\d_{u_2v_3}-\d_{u_1v_4}\d_{v_1v_3}\d_{u_2v_2})
                              \nn\\
 &&+\d_{u_1v_1}\d_{u_2v_2}\d_{u_3v_3}\d_{u_4v_4}
   +\d_{u_1v_1}\d_{u_2v_2}\d_{u_3v_4}\d_{u_4v_3}
   +\d_{u_1v_1}\d_{u_2v_3}\d_{u_3v_2}\d_{u_4v_4}\nn\\
  && -\d_{u_1v_1}\d_{u_2v_3}\d_{u_4v_2}\d_{u_3v_4}
     +\d_{u_1v_1}\d_{u_2v_4}\d_{u_3v_2}\d_{u_4v_3}
     +\d_{u_1v_1}\d_{u_2v_4}\d_{u_4v_2}\d_{u_3v_3}
     \nn\\
  &&+\d_{u_1v_2}\d_{u_2v_1}\d_{u_3v_3}\d_{u_4v_4}
    +\d_{u_1v_2}\d_{u_2v_1}\d_{u_3v_4}\d_{u_4v_3}
    -\d_{u_1v_2}\d_{u_3v_1}\d_{u_2v_3}\d_{u_4v_4}\nn\\
  &&-\d_{u_1v_2}\d_{u_3v_1}\d_{u_2v_4}\d_{u_4v_3}
    +\d_{u_1v_2}\d_{u_4v_1}\d_{u_2v_3}\d_{u_3v_4}
    -\d_{u_1v_2}\d_{u_4v_1}\d_{u_2v_4}\d_{u_3v_3}
    \nn\\
  &&+\d_{u_1v_3}\d_{u_2v_1}\d_{u_3v_2}\d_{u_4v_4}
    -\d_{u_1v_3}\d_{u_2v_1}\d_{u_4v_2}\d_{u_3v_4}
    +\d_{u_1v_3}\d_{u_3v_1}\d_{u_2v_2}\d_{u_4v_4}\nn\\
  &&+\d_{u_1v_3}\d_{u_3v_1}\d_{u_2v_4}\d_{u_4v_2}
    -\d_{u_1v_3}\d_{u_4v_1}\d_{u_2v_2}\d_{u_3v_4}
    -\d_{u_1v_3}\d_{u_4v_1}\d_{u_2v_4}\d_{u_3v_2}
    \nn\\
  &&+\d_{u_1v_4}\d_{u_2v_1}\d_{u_3v_2}\d_{u_4v_3}
    +\d_{u_1v_4}\d_{u_2v_1}\d_{u_4v_2}\d_{u_3v_3}
    +\d_{u_1v_4}\d_{u_3v_1}\d_{u_2v_2}\d_{u_4v_3}\nn\\
  &&-\d_{u_1v_4}\d_{u_3v_1}\d_{u_2v_3}\d_{u_4v_2}
    +\d_{u_1v_4}\d_{u_4v_1}\d_{u_2v_2}\d_{u_3v_3}
    +\d_{u_1v_4}\d_{u_4v_1}\d_{u_2v_3}\d_{u_3v_2}
 \eea
The following three relations are also used in sec3.
 \bea
 &&<S(x_1)\g^{A_1B_1} S(x_1)\;S(x_2)\g^{A_2B_2} S(x_2)
 \;S(y)\g^{A_3B_3} S(y)
 \;S(y)\g^{A_4B_4} S(y)>\nn\\
 =&&
 \frac{4 \Tr[{\g^{A_1B_1}}{\g^{A_4B_4}}] \Tr[{\g^{A_2B_2}}{\g^{A_3B_3}}]+4 \Tr[{\g^{A_1B_1}}{\g^{A_3B_3}}]
  \Tr[{\g^{A_2B_2}}{\g^{A_4B_4}}]-16 \Tr[{\g^{A_1B_1}}{\g^{A_3B_3}}{\g^{A_2B_2}}{\g^{A_4B_4}}]}{(x_1-y)^2 (x_2-y)^2}
 \nn\\
 \eea

 \bea
 &&<S(x_1)\g^{A_1B_1} S(x_1)\;S(x_2)\g^{A_2B_2} S(x_2)
 \;S(x_3)\g^{A_3B_3} S(x_3)\;S(y)\g^{A_4B_4} S(y)
 \;S(y)\g^{A_5B_5} S(y)>\nn\\
 =&&
 +\frac{16\Tr[\g^{A_3B_3}\g^{A_5B_5}] \Tr[\g^{A_1B_1}\g^{A_2B_2}\g^{A_4B_4}]}{(x_1-x_2)
   (x_1-y) (x_2-y) (x_3-y)^2}+\frac{16 \Tr[\g^{A_3B_3}\g^{A_4B_4}]\Tr[\g^{A_1B_1}\g^{A_2B_2}\g^{A_5B_5}]}
   {(x_1-x_2) (x_3-y)^2 (x_1-y)(x_2-y)}\nn\\
   &&
   +\frac{16 \Tr[\g^{A_2B_2}\g^{A_5B_5}] \Tr[\g^{A_1B_1}\g^{A_3B_3}{\g^{A_4B_4}}]}
   {(x_1-x_3) (x_1-y) (x_3-y) (x_2-y)^2}+\frac{16\Tr[\g^{A_2B_2}\g^{A_4B_4}] \Tr[\g^{A_1B_1}\g^{A_3B_3}\g^{A_5B_5}]}
   {(x_1-x_3) (x_2-y)^2(x_1-y) (x_3-y)}\nn\\
   &&
   \nn\\
   &&+\frac{16 \Tr[\g^{A_1B_1}\g^{A_5B_5}]\Tr[\g^{A_2B_2}\g^{A_3B_3}\g^{A_4B_4}]}
   {(x_2-x_3) (x_2-y)(x_3-y) (x_1-y)^2}
   +\frac{16 \Tr[\g^{A_1B_1}\g^{A_4B_4}] \Tr[\g^{A_2B_2}\g^{A_3B_3}\g^{A_5B_5}]}
   {(x_2-x_3) (x_1-y)^2 (x_2-y) (x_3-y)}\nn\\
   &&
   \nn\\
   &&
   \nn\\
  && -\frac{32\Tr[\g^{A_1B_1}\g^{A_2B_2}\g^{A_4B_4}\g^{A_3B_3}\g^{A_5B_5}]}
   {(x_1-x_2)(x_2-y) (x_3-y) (x_1-y) (x_3-y)}
   -\frac{32 \Tr[\g^{A_1B_1}\g^{A_2B_2}\g^{A_5B_5}\g^{A_3B_3}\g^{A_4B_4}]}
   {(x_1-x_2) (x_1-y) (x_3-y) (x_2-y)(x_3-y)}\nn\\
    &&-\frac{32 \Tr[\g^{A_1B_1}\g^{A_3B_3}\g^{A_4B_4}\g^{A_2B_2}\g^{A_5B_5}]}
   {(x_1-x_3) (x_2-y) (x_3-y) (x_1-y)(x_2-y)}
  -\frac{32 \Tr[\g^{A_1B_1}\g^{A_3B_3}\g^{A_5B_5}\g^{A_2B_2}\g^{A_4B_4}]}
   {(x_1-x_3)(x_1-y) (x_2-y) (x_2-y) (x_3-y)}\nn\\
   &&-\frac{32\Tr[\g^{A_1B_1}\g^{A_4B_4}\g^{A_2B_2}\g^{A_3B_3}\g^{A_5B_5}]}
   {(x_2-x_3)(x_1-y) (x_2-y) (x_1-y) (x_3-y)}
   +\frac{32 \Tr[\g^{A_1B_1}\g 4\g^{A_3B_3}\g^{A_2B_2}\g^{A_5B_5}]}
   {(x_2-x_3) (x_1-y) (x_3-y)(x_1-y) (x_2-y)}\nn\\
 \eea
and
 \bea
 &&<S(x_1)\g^{A_1B_1} S(x_1)\;S(x_2)\g^{A_2B_2} S(x_2)
 \;S(x_3)\g^{A_3B_3} S(x_3)\;S(x_4)\g^{A_4B_4} S(x_4)\nn\\
 &&\hspace{1in}\;S(y)\g^{A_5B_5} S(y)\;S(y)\g^{A_6B_6} S(y)>\nn\\
 =&&\frac{8\Tr[{\g^{A_3B_3}}{\g^{A_4B_4}}]}{(x_3-x_4)^2 (x_2-y)^2}
 ( \Tr[{\g^{A_1B_1}}{\g^{A_6B_6}}]\Tr[{\g^{A_2B_2}}{\g^{A_5B_5}}]
 + \Tr[{\g^{A_1B_1}}{\g^{A_5B_5}}]
 \Tr[{\g^{A_2B_2}}{\g^{A_6B_6}}]\nn\\
  &&\hspace{3in}
  -4 \Tr[{\g^{A_1B_1}}{\g^{A_5B_5}}{\g^{A_2B_2}}{\g^{A_6B_6}}])\nn\\
  && +\frac{8\Tr[{\g^{A_2B_2}}{\g^{A_4B_4}}] } {(x_2-x_4)^2 (x_3-y)^2}
      ( \Tr[{\g^{A_1B_1}}{\g^{A_6B_6}}] \Tr[{\g^{A_3B_3}}{\g^{A_5B_5}}]
   + \Tr[{\g^{A_1B_1}}{\g^{A_5B_5}}]
   \Tr[{\g^{A_3B_3}}{\g^{A_6B_6}}]\nn\\
  && \hspace{3in}-4 \Tr[{\g^{A_1B_1}}{\g^{A_5B_5}}{\g^{A_3B_3}}{\g^{A_6B_6}}])\nn\\
  &&+\frac{8\Tr[{\g^{A_2B_2}}{\g^{A_3B_3}}]}{(x_2-x_3)^2 (x_4-y)^2}
  ( \Tr[{\g^{A_1B_1}}{\g^{A_6B_6}}]\Tr[{\g^{A_4B_4}}{\g^{A_5B_5}}]
  +\Tr[{\g^{A_1B_1}}{\g^{A_5B_5}}]\Tr[{\g^{A_4B_4}}{\g^{A_6B_6}}]\nn\\
   &&\hspace{3in}-4\Tr[{\g^{A_1B_1}}{\g^{A_5B_5}}{\g^{A_4B_4}}{\g^{A_6B_6}}])
    \nn\\
  && +\frac{32}{(x_2-x_4) (x_3-x_4) (x_2-y) (x_3-y)}
   (- \Tr[{\g^{A_1B_1}}{\g^{A_6B_6}}] \Tr[{\g^{A_2B_2}}{\g^{A_4B_4}}{\g^{A_3B_3}}{\g^{A_5B_5}}]
   \nn\\
   &&\hspace{2.7in}-\Tr[{\g^{A_1B_1}}{\g^{A_5B_5}}] \Tr[{\g^{A_2B_2}}
            {\g^{A_4B_4}}{\g^{A_3B_3}}{\g^{A_6B_6}}]\nn\\
   &&\hspace{3in}+2\Tr[{\g^{A_1B_1}}{\g^{A_5B_5}}{\g^{A_2B_2}}
      {\g^{A_4B_4}}{\g^{A_3B_3}}{\g^{A_6B_6}}]\nn\\
   &&\hspace{3in}+2\Tr[{\g^{A_1B_1}}{\g^{A_5B_5}}{\g^{A_3B_3}}
        {\g^{A_4B_4}}{\g^{A_2B_2}}{\g^{A_6B_6}}])\nn\\
  && +\frac{32}{(x_2-x_3) (x_3-x_4) (x_2-y)(x_4-y)}
  ( \Tr[{\g^{A_1B_1}}{\g^{A_6B_6}}]\Tr[{\g^{A_2B_2}}{\g^{A_3B_3}}{\g^{A_4B_4}}{\g^{A_5B_5}}]\nn\\
  &&\hspace{2.7in}+ \Tr[{\g^{A_1B_1}}{\g^{A_5B_5}}]
          \Tr[{\g^{A_2B_2}}{\g^{A_3B_3}}{\g^{A_4B_4}}{\g^{A_6B_6}}]\nn\\
  && \hspace{3in}-2 \Tr[{\g^{A_1B_1}}{\g^{A_5B_5}}{\g^{A_2B_2}}
               {\g^{A_3B_3}}{\g^{A_4B_4}}{\g^{A_6B_6}}]\nn\\
   &&\hspace{3in}-2 \Tr[{\g^{A_1B_1}}{\g^{A_5B_5}}{\g^{A_4B_4}}{\g^{A_3B_3}}{\g^{A_2B_2}}{\g^{A_6B_6}}])\nn\\
   &&+\frac{32}{(x_2-x_3) (x_2-x_4) (x_3-y) (x_4-y)}
  (- \Tr[{\g^{A_1B_1}}{\g^{A_6B_6}}] \Tr[{\g^{A_2B_2}}{\g^{A_3B_3}}{\g^{A_5B_5}}{\g^{A_4B_4}}]\nn\\
   &&\hspace{2.7in}- \Tr[{\g^{A_1B_1}}{\g^{A_5B_5}}] \Tr[{\g^{A_2B_2}}
              {\g^{A_3B_3}}{\g^{A_6B_6}}{\g^{A_4B_4}}]\nn\\
  && \hspace{3in}+2 \Tr[{\g^{A_1B_1}}{\g^{A_6B_6}}{\g^{A_3B_3}}
                 {\g^{A_2B_2}}{\g^{A_4B_4}}{\g^{A_5B_5}}]\nn\\
  &&\hspace{3in} +2 \Tr[{\g^{A_1B_1}}{\g^{A_6B_6}}{\g^{A_4B_4}}{\g^{A_2B_2}}{\g^{A_3B_3}}{\g^{A_5B_5}}])
   \nn\\
 \eea

\renewcommand{\theequation}{C.\arabic{equation}}
\setcounter{equation}{0}

\section{More details on proof of (\ref{4szero})}

In sec3, it was stated that the four $S$-correlator eq(\ref{4szero})
vanishes. As an illustration, the computation for the coefficient of
$(\z_2\cdot \z_3)$ was presented. Below we give two additional
coefficients, the coefficient of $(\z_2\cdot \z_4)$ and the
coefficient of $\z\cdot k\;\z\cdot k\;\z\cdot k\;\z\cdot k$.

\subsection{coeff. of $(\z_2\cdot \z_4)$}

In sec3, it was stated
 \bea
 &&<V^{u_1}_v(x_1) V^{u_2}_v(x_2) V^{u_3}_v(x_3) V^{u_4}_v(x_4)
 ;S\g^{AB}SS\g^{CD}S>=0
 \eea
We showed that terms with $(\z_2\cdot \z_3)$ add to zero. Here we
present more details taking terms with a factor, $\z_2\cdot \z_4$,
and terms with a factor, $\z\cdot k\;\z\cdot k\;\z\cdot k\;\z\cdot
k$ for illustrations. One can show that
 \bea
 && -\fr{17}{96}<VVVV\;(S\g^{uv} S)( S\g^{uv} S)>\nn\\
 =&&-\fr{17}{3}\fr{l^4}{\e x_1^2} \z_2\cdot \z_4\left\{\fr{}{}\right.4
    \left[\frac{  \alpha
   '}{(x_2-x_4)^2}\right] \z_1\cdot k_3\;\z_3\cdot k_1\nn\\
 &&-{l^2}\left[\fr{}{}\right.
   \frac{(-2 \alpha ') } { (x_2-x_4) }
   \left(\frac{i \z_3\cdot k_2}{x_3-x_2}+\frac{i \z_3\cdot k_4}{x_3-x_4}\right)
   (\z_1\cdot k_2 t-\z_1\cdot k_4 s)
 \left.\fr{}{}\right]\nn\\
 &&-{l^4}
 \left[\fr{}{}\right.
   \frac{1}{{\left( x_2 - x_4 \right) }^2\,}
   (\z_1\cdot k_3\;\z_3\cdot k_1)u
        \nn\\
  &&+ \frac{1}{2(x_3 -x_4)\left( x_2 -x_4 \right) \, }
 (-s\z_1\cdot k_2\;\z_3\cdot k_1+u\z_1\cdot k_2\;\z_3\cdot k_4
 +t\z_1\cdot k_4\;\z_3\cdot k_1\nn\\
 &&\hspace{.4in}-u\z_1\cdot k_4\;\z_3\cdot k_2-s\z_1\cdot k_3\;\z_3\cdot k_4
  +t\z_1\cdot k_3\;\z_3\cdot k_2-u\z_1\cdot k_3\;\z_3\cdot k_1)
          \nn\\
  && - \frac{1}{2(x_3 -x_4)\left( x_2 -x_3 \right) \, }
 (s\z_1\cdot k_2\;\z_3\cdot k_1-u\z_1\cdot k_2\;\z_3\cdot k_4
 +s\z_1\cdot k_3\;\z_3\cdot k_4\nn\\
  &&\hspace{.4in}-u\z_1\cdot k_3\;\z_3\cdot k_1
  +t\z_1\cdot k_3\;\z_3\cdot k_2
  +t\z_1\cdot k_4\;\z_3\cdot k_1-u\z_1\cdot k_4\;\z_3\cdot k_2) \nn\\
     &&+\frac{1}{2\left( x_2 -x_3 \right) \,\left( x_2 - x_4 \right) \, }
 (-u\z_1\cdot k_3\;\z_3\cdot k_1+s\z_1\cdot k_3\;\z_3\cdot k_4
         -t\z_1\cdot k_3\;\z_3\cdot k_2\nn\\
  &&\hspace{.4in}
 +s\z_1\cdot k_2\;\z_3\cdot k_1-u\z_1\cdot k_2\;\z_3\cdot k_4
 -t\z_1\cdot k_4\;\z_3\cdot k_1
  +u\z_1\cdot k_4\;\z_3\cdot k_2)
       \left.\fr{}{}\right]\left.\fr{}{}\right\}\nn\\
 \eea
Similarly,
 \bea
 && -\fr{1}{6}<VVVV\;(S\g^{ab} S)( S\g^{ab} S)>\nn\\
  =&&\fr{112}{3}
  \fr{l^4}{\e x_1^2}\;\z_2\cdot \z_4
  \left\{\fr{}{}\right.
  4\left[\frac{  \alpha
   '}{(x_2-x_4)^2}\right]  \z_1\cdot k_3\;\z_3\cdot k_1\nn\\
   &&-l^2\left[\fr{}{}\right.
    (-2 \alpha ')
   \left(\frac{i \z_3\cdot k_2}{x_3-x_2}+\frac{i \z_3\cdot k_4}{x_3-x_4}\right)
      \frac{1 } { (x_2-x_4) }
      (\z_1\cdot k_2\;t-\z_1\cdot k_4 s)
 \left.\fr{}{}\right]\nn\\
 &&-{l^4}
 \left[\fr{}{}\right.
   \frac{1}{{\left( x_2 - x_4 \right) }^2\,}
   (\z_1\cdot k_3\;\z_3\cdot k_1)u
        \nn\\
  &&+ \frac{1}{2(x_3 -x_4)\left( x_2 -x_4 \right) \, }
 (-s\z_1\cdot k_2\;\z_3\cdot k_1+u\z_1\cdot k_2\;\z_3\cdot k_4
 +t\z_1\cdot k_4\;\z_3\cdot k_1\nn\\
 &&\hspace{.4in}-u\z_1\cdot k_4\;\z_3\cdot k_2-s\z_1\cdot k_3\;\z_3\cdot k_4
  +t\z_1\cdot k_3\;\z_3\cdot k_2-u\z_1\cdot k_3\;\z_3\cdot k_1)
          \nn\\
  && - \frac{1}{2(x_3 -x_4)\left( x_2 -x_3 \right) \, }
 (s\z_1\cdot k_2\;\z_3\cdot k_1-u\z_1\cdot k_2\;\z_3\cdot k_4
 +s\z_1\cdot k_3\;\z_3\cdot k_4\nn\\
  &&\hspace{.4in}-u\z_1\cdot k_3\;\z_3\cdot k_1
  +t\z_1\cdot k_3\;\z_3\cdot k_2
  +t\z_1\cdot k_4\;\z_3\cdot k_1-u\z_1\cdot k_4\;\z_3\cdot k_2) \nn\\
     &&+\frac{1}{2\left( x_2 -x_3 \right) \,\left( x_2 - x_4 \right) \, }
 (-u\z_1\cdot k_3\;\z_3\cdot k_1
 +s\z_1\cdot k_3\;\z_3\cdot k_4-t\z_1\cdot k_3\;\z_3\cdot k_2\nn\\
  &&\hspace{.4in}
 +s\z_1\cdot k_2\;\z_3\cdot k_1-u\z_1\cdot k_2\;\z_3\cdot k_4
 -t\z_1\cdot k_4\;\z_3\cdot k_1
  +u\z_1\cdot k_4\;\z_3\cdot k_2)
       \left.\fr{}{}\right]\left.\fr{}{}\right\}\nn\\
 \eea
Now consider the terms that come with $\fr{q^2}{r_0^4}X_0^aX_0^b$:
 \bea
&& \fr{35}{24}<VVVV\;(S\g^{ac} S)( S\g^{bc} S)>\nn\\
=&& -\fr{35\cdot 7\cdot 4}{24} \left\{\fr{}{}\right.
  \fr{l^4}{\e x_1^2}\;\z_2\cdot \z_4
  4\left[\frac{  \alpha
   '}{(x_2-x_4)^2}\right]  \z_1\cdot k_3\;\z_3\cdot k_1\nn\\
   &&-l^2\left[\fr{}{}\right.
    (-2 \alpha ')
   \left(\frac{i \z_3\cdot k_2}{x_3-x_2}+\frac{i \z_3\cdot k_4}{x_3-x_4}\right)
      \frac{1 } { (x_2-x_4) }
      (\z_1\cdot k_2\;t-\z_1\cdot k_4 s)
 \left.\fr{}{}\right]\nn\\
 &&-{l^4}
 \left[\fr{}{}\right.
   \frac{1}{{\left( x_2 - x_4 \right) }^2\,}
   (\z_1\cdot k_3\;\z_3\cdot k_1)u
        \nn\\
  &&+ \frac{1}{2(x_3 -x_4)\left( x_2 -x_4 \right) \, }
 (-s\z_1\cdot k_2\;\z_3\cdot k_1+u\z_1\cdot k_2\;\z_3\cdot k_4
 +t\z_1\cdot k_4\;\z_3\cdot k_1\nn\\
 &&\hspace{.4in}-u\z_1\cdot k_4\;\z_3\cdot k_2-s\z_1\cdot k_3\;\z_3\cdot k_4
  +t\z_1\cdot k_3\;\z_3\cdot k_2-u\z_1\cdot k_3\;\z_3\cdot k_1)
          \nn\\
  && - \frac{1}{2(x_3 -x_4)\left( x_2 -x_3 \right) \, }
 (s\z_1\cdot k_2\;\z_3\cdot k_1-u\z_1\cdot k_2\;\z_3\cdot k_4
 +s\z_1\cdot k_3\;\z_3\cdot k_4\nn\\
  &&\hspace{.4in}-u\z_1\cdot k_3\;\z_3\cdot k_1
  +t\z_1\cdot k_3\;\z_3\cdot k_2
  +t\z_1\cdot k_4\;\z_3\cdot k_1-u\z_1\cdot k_4\;\z_3\cdot k_2) \nn\\
     &&+\frac{1}{2\left( x_2 -x_3 \right) \,\left( x_2 - x_4 \right) \, }
 (-u\z_1\cdot k_3\;\z_3\cdot k_1+s\z_1\cdot k_3\;\z_3\cdot k_4
 -t\z_1\cdot k_3\;\z_3\cdot k_2\nn\\
  &&\hspace{.4in}
 +s\z_1\cdot k_2\;\z_3\cdot k_1-u\z_1\cdot k_2\;\z_3\cdot k_4
 -t\z_1\cdot k_4\;\z_3\cdot k_1
  +u\z_1\cdot k_4\;\z_3\cdot k_2)
       \left.\fr{}{}\right]\left.\fr{}{}\right\}\nn\\
 \eea
Therefore apart from some common factors one gets after taking Wick
rotation into account
 \bea
   &&  \left\{ \left( \fr{q^2}{r_0^2}\right)
  \left[-\fr{17}{96}(S\g^{uv} S)( S\g^{uv} S)
  -\fr16(S\g^{ab} S)( S\g^{ab} S)\right]\right.
   \nn\\
  &&\left.+\left(\fr{q^2}{r_0^4}\right) X_0^aX_0^b
  \left[\fr{35}{24}(S\g^{ac} S)( S\g^{bc} S)
  \right]\right\}\nn\\
\Rightarrow &&
  -\frac{  2}{(x_2-x_4)^2}  \z_1\cdot k_3\;\z_3\cdot k_1\nn\\
   &&- \left(\frac{ \z_3\cdot k_2}{x_3-x_2}+\frac{ \z_3\cdot k_4}{x_3-x_4}\right)
      \frac{1 } { (x_2-x_4) }
      (\z_1\cdot k_2\;t-\z_1\cdot k_4 s)
 \nn\\
 &&-
 \left[\fr{}{}\right.
   \frac{1}{{\left( x_2 - x_4 \right) }^2\,}
   (\z_1\cdot k_3\;\z_3\cdot k_1)u
        \nn\\
  &&+ \frac{1}{2(x_3 -x_4)\left( x_2 -x_4 \right) \, }
 (-s\z_1\cdot k_2\;\z_3\cdot k_1+u\z_1\cdot k_2\;\z_3\cdot k_4
 +t\z_1\cdot k_4\;\z_3\cdot k_1\nn\\
 &&\hspace{.4in}-u\z_1\cdot k_4\;\z_3\cdot k_2-s\z_1\cdot k_3\;\z_3\cdot k_4
  +t\z_1\cdot k_3\;\z_3\cdot k_2-u\z_1\cdot k_3\;\z_3\cdot k_1)
          \nn\\
  && - \frac{1}{2(x_3 -x_4)\left( x_2 -x_3 \right) \, }
 (s\z_1\cdot k_2\;\z_3\cdot k_1-u\z_1\cdot k_2\;\z_3\cdot k_4
 +s\z_1\cdot k_3\;\z_3\cdot k_4\nn\\
  &&\hspace{.4in}-u\z_1\cdot k_3\;\z_3\cdot k_1
  +t\z_1\cdot k_3\;\z_3\cdot k_2
  +t\z_1\cdot k_4\;\z_3\cdot k_1-u\z_1\cdot k_4\;\z_3\cdot k_2) \nn\\
     &&+\frac{1}{2\left( x_2 -x_3 \right) \,\left( x_2 - x_4 \right) \, }
 (-u\z_1\cdot k_3\;\z_3\cdot k_1
 +s\z_1\cdot k_3\;\z_3\cdot k_4-t\z_1\cdot k_3\;\z_3\cdot k_2\nn\\
  &&\hspace{.4in}
 +s\z_1\cdot k_2\;\z_3\cdot k_1-u\z_1\cdot k_2\;\z_3\cdot k_4
 -t\z_1\cdot k_4\;\z_3\cdot k_1
  +u\z_1\cdot k_4\;\z_3\cdot k_2)
       \left.\fr{}{}\right]\nn\\
 =&& -2  \z_1\cdot k_3\;\z_3\cdot k_1\nn\\
   &&- \left(\frac{ \z_3\cdot k_2}{x-1}+\frac{ \z_3\cdot k_4}{x}\right)
      (\z_1\cdot k_2\;t-\z_1\cdot k_4 s)
 \nn\\
 &&
   -(\z_1\cdot k_3\;\z_3\cdot k_1)u
        \nn\\
  &&- \frac{1}{2x }
 (-s\z_1\cdot k_2\;\z_3\cdot k_1+u\z_1\cdot k_2\;\z_3\cdot k_4
 +t\z_1\cdot k_4\;\z_3\cdot k_1\nn\\
 &&\hspace{.4in}-u\z_1\cdot k_4\;\z_3\cdot k_2-s\z_1\cdot k_3\;\z_3\cdot k_4
  +t\z_1\cdot k_3\;\z_3\cdot k_2-u\z_1\cdot k_3\;\z_3\cdot k_1)
          \nn\\
  && + \frac{1}{2x\left( 1 -x \right) \, }
 (s\z_1\cdot k_2\;\z_3\cdot k_1-u\z_1\cdot k_2\;\z_3\cdot k_4
 +s\z_1\cdot k_3\;\z_3\cdot k_4\nn\\
  &&\hspace{.4in}-u\z_1\cdot k_3\;\z_3\cdot k_1
  +t\z_1\cdot k_3\;\z_3\cdot k_2
  +t\z_1\cdot k_4\;\z_3\cdot k_1-u\z_1\cdot k_4\;\z_3\cdot k_2) \nn\\
     &&-\frac{1}{2\left( 1 -x \right)  }
 (-u\z_1\cdot k_3\;\z_3\cdot k_1
 +s\z_1\cdot k_3\;\z_3\cdot k_4-t\z_1\cdot k_3\;\z_3\cdot k_2\nn\\
  &&\hspace{.4in}
 +s\z_1\cdot k_2\;\z_3\cdot k_1-u\z_1\cdot k_2\;\z_3\cdot k_4
 -t\z_1\cdot k_4\;\z_3\cdot k_1
  +u\z_1\cdot k_4\;\z_3\cdot k_2)
 \eea
which leads to a vanishing result after the $x$-integration.

\subsection{coeff. of $\z\cdot k\;\z\cdot k\;\z\cdot k\;\z\cdot
k\;$}

Similarly one finds after some algebra
 \bea
   &&  \left\{ \left( \fr{q^2}{r_0^2}\right)
  \left[-\fr{17}{96}(S\g^{uv} S)( S\g^{uv} S)
  -\fr16(S\g^{ab} S)( S\g^{ab} S)\right]\right.
   \nn\\
  &&\left.+\left(\fr{q^2}{r_0^4}\right) X_0^aX_0^b
  \left[\fr{35}{24}(S\g^{ac} S)( S\g^{bc} S)
  \right]\right\}\nn\\
= && \quad-\left(-{ \z_4\cdot k_2}
    -\frac{ \z_4\cdot k_3}{x}\right) \left(\frac{
   \z_3\cdot k_2}{x-1}+\frac{ \z_3\cdot k_4}{x}\right)
     \z_1\cdot k_2\;\z_2\cdot k_1\nn\\
  &&\quad\;\;-
   \left(\frac{ \z_2\cdot k_3}{1-x}
   +{ \z_2\cdot k_4}\right) \left(\frac{
   \z_3\cdot k_2}{x-1}+\frac{\z_3\cdot k_4}{x}\right)
    \z_1\cdot k_4\;\z_4\cdot k_1\nn\\
   &&\quad\;\;-
    \left(-{ \z_4\cdot k_2}
    -\frac{ \z_4\cdot k_3}{x}\right)
     \left(\frac{\z_2\cdot k_3}{1-x}+{ \z_2\cdot k_4}\right)
   \z_1\cdot k_3\;\z_3\cdot k_1
      \nn\\
 &&+
   \left(\frac{ \z_2\cdot k_3}{1-x}
             +{ \z_2\cdot k_4}\right)
    \frac{ 1 } { x }
    (-\z_3\cdot k_4\;\z_1\cdot k_3\;\z_4\cdot k_1
           +\z_4\cdot k_3\;\z_1\cdot k_4\;\z_3\cdot k_1)\nn\\
 &&\quad\;\;+
   \left(\frac{ \z_3\cdot k_2}{x-1}
             +\frac{ \z_3\cdot k_4}{x}\right)
       (-\z_2\cdot k_4\;\z_1\cdot k_2\;\z_4\cdot k_1
           +\z_4\cdot k_2\;\z_1\cdot k_4\;\z_2\cdot k_1)\nn\\
 &&\quad\;\;+
   \left(-{ \z_4\cdot k_2}
           -\frac{ \z_4\cdot k_3}{x}\right)
       \frac{1} { (1-x) }
       (-\z_2\cdot k_3\;\z_1\cdot k_2\;\z_3\cdot k_1
           +\z_3\cdot k_2\;\z_1\cdot k_3\;\z_2\cdot k_1)
  \nn\\
 &&
  -\frac{1}{{x }^2\,}
   \z_1\cdot k_2\;\z_2\cdot k_1\;\z_3\cdot k_4\;\z_4\cdot k_3
        -\z_1\cdot k_3\;\z_3\cdot k_1\;\z_2\cdot k_4\;\z_4\cdot k_2
    \nn\\
  && \quad -\frac{1}{{\left( 1 - x \right) }^2\,}
    \z_1\cdot k_4\;\z_4\cdot k_1\;\z_2\cdot k_3\;\z_3\cdot k_2
       \nn\\
  && +\frac{1}{2x  }
        \nn\\
&&(\z_1\cdot k_2\;\z_2\cdot k_1\;\z_4\cdot k_3\;\z_3\cdot k_4
   -\z_1\cdot k_2\;\z_4\cdot k_1\;\z_2\cdot k_3\;\z_3\cdot k_4
  +\z_1\cdot k_2\;\z_3\cdot k_1\;\z_2\cdot k_4\;\z_4\cdot k_3
   \nn\\
  && -\z_1\cdot k_4\;\z_2\cdot k_1\;\z_3\cdot k_2\;\z_4\cdot k_3
   +\z_1\cdot k_4\;\z_4\cdot k_1\;\z_2\cdot k_3\;\z_3\cdot k_2
   -\z_1\cdot k_4\;\z_3\cdot k_1\;\z_2\cdot k_3\;\z_4\cdot k_2
    \nn\\
  &&+\z_1\cdot k_3\;\z_2\cdot k_1\;\z_4\cdot k_2\;\z_3\cdot k_4
    -\z_1\cdot k_3\;\z_4\cdot k_1\;\z_2\cdot k_4\;\z_3\cdot k_2
    +\z_1\cdot k_3\;\z_3\cdot k_1\;\z_2\cdot k_4\;\z_4\cdot k_2 )
          \nn\\
  && -\frac{1}{2x \left( 1 -x \right) \, }\nn\\
  &&(\z_1\cdot k_2\;\z_2\cdot k_1\;\z_3\cdot k_4\;\z_4\cdot k_3
    -\z_1\cdot k_2\;\z_3\cdot k_1\;\z_2\cdot k_4\;\z_4\cdot k_3
    +\z_1\cdot k_2\;\z_4\cdot k_1\;\z_2\cdot k_3\;\z_3\cdot k_4
   \nn\\
  &&-\z_1\cdot k_3\;\z_2\cdot k_1\;\z_4\cdot k_2\;\z_3\cdot k_4
    +\z_1\cdot k_3\;\z_3\cdot k_1\;\z_2\cdot k_4\;\z_4\cdot k_2
    -\z_1\cdot k_3\;\z_4\cdot k_1\;\z_2\cdot k_4\;\z_3\cdot k_2
    \nn\\
  && \z_1\cdot k_4\;\z_2\cdot k_1\;\z_3\cdot k_2\;\z_4\cdot k_3
    -\z_1\cdot k_4\;\z_3\cdot k_1\;\z_2\cdot k_3\;\z_4\cdot k_2
    +\z_1\cdot k_4\;\z_4\cdot k_1;\z_2\cdot k_3\;\z_3\cdot k_2)
    \nn\\
     &&+\frac{1}{2\left( 1 -x \right) }
      \nn\\
  &&(\z_1\cdot k_3\;\z_3\cdot k_1\;\z_2\cdot k_4\;\z_4\cdot k_2
    -\z_1\cdot k_3\;\z_2\cdot k_1\;\z_3\cdot k_4\;\z_4\cdot k_2
   +\z_1\cdot k_3\;\z_4\cdot k_1\;\z_3\cdot k_2\;\z_2\cdot k_4
   \nn\\
  &&-\z_1\cdot k_2\;\z_3\cdot k_1\;\z_4\cdot k_3\;\z_2\cdot k_4
    +\z_1\cdot k_2\;\z_2\cdot k_1\;\z_3\cdot k_4\;\z_4\cdot k_3
   -\z_1\cdot k_2\;\z_4\cdot k_1\;\z_3\cdot k_4\;\z_2\cdot k_3
    \nn\\
  && +\z_1\cdot k_4\;\z_3\cdot k_1\;\z_2\cdot k_3\;\z_4\cdot k_2
    -\z_1\cdot k_4\;\z_2\cdot k_1\;\z_3\cdot k_2\;\z_4\cdot k_3
   +\z_1\cdot k_4\;\z_4\cdot k_1\;\z_3\cdot k_2\;\z_2\cdot k_3)
       \nn\\
 \eea
 where Wick rotation has been taken into account.
Once again the $x$-integration yields zero.

\newpage

\end{document}